\documentclass[11pt]{article}

\usepackage[final]{naacl_2024}
\usepackage[T1]{fontenc}

\usepackage{times}
\usepackage{latexsym}
\usepackage{url}            % simple URL typesetting
\usepackage{booktabs}       % professional-quality tables
\usepackage{amsfonts}       % blackboard math symbols
\usepackage{nicefrac}       % compact symbols for 1/2, etc.
\usepackage{microtype}      % microtypography
\usepackage{xcolor}         % colors
\usepackage{color, colortbl}
\usepackage{graphicx}
\usepackage{wrapfig,lipsum}
\usepackage{floatrow}
\usepackage{amssymb}
\usepackage{amsmath}
\usepackage{paralist}
\usepackage{pifont}% http://ctan.org/pkg/pifont
\usepackage{xcolor,colortbl}
\usepackage{caption}
\usepackage{enumitem}
\usepackage{pdfpages}
\usepackage[hang,flushmargin]{footmisc}
\definecolor{dark_green}{rgb}{0.0, 0.5, 0.0}
\definecolor{coralpink}{rgb}{0.97, 0.51, 0.47}
\definecolor{lightpink}{rgb}{1.0, 0.71, 0.76}
\definecolor{paleaqua}{rgb}{0.74, 0.83, 0.9}

\newcommand{\blue}[1]{\textcolor{blue}{#1}}
\newcommand{\red}[1]{\textcolor{red}{#1}}
\newcommand{\green}[1]{\textcolor{dark_green}{#1}}

\usepackage[T1]{fontenc}
\usepackage[utf8]{inputenc}

\usepackage{microtype}

\usepackage{inconsolata}

\title{\vspace{-10mm}\includegraphics[trim=0 0 0 0,clip,scale=0.25]{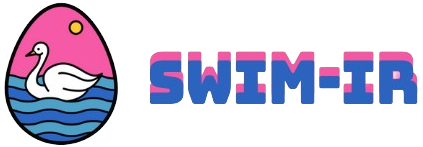} \\ Leveraging LLMs for Synthesizing Training Data Across Many Languages in Multilingual Dense Retrieval}

\author{Nandan Thakur\thanks{$^*$Work done while Nandan was a student researcher at Google Research.~$^\dagger$Correspondence to: Nandan Thakur <nandan.thakur@uwaterloo.ca>, Jianmo Ni <jianmon@google.com>, Daniel Cer <cer@google.com>.}~\,$^\dagger$\blue{$^\mathsection$}, Jianmo Ni$^\dagger$\green{$^\heartsuit$}, Gustavo Hern\'andez \'Abrego\red{$^\lozenge$} \\
\textbf{John Wieting}\green{$^\heartsuit$}, \textbf{Jimmy Lin}\blue{$^\mathsection$}, \textbf{Daniel Cer}$^\dagger$\red{$^\lozenge$}\\ \\ 
  \red{$^\lozenge$}Google Research, \green{$^\heartsuit$}Google DeepMind, \blue{$^\mathsection$}University of Waterloo\\ \\
}

\begin{document}

\maketitle

\begin{abstract}
There has been limited success for dense retrieval models in multilingual retrieval, due to uneven and scarce training data available across multiple languages.
Synthetic training data generation is promising~(e.g., InPars or Promptagator), but has been investigated only for English. Therefore, to study model capabilities across both cross-lingual and monolingual retrieval tasks, we develop \textbf{SWIM-IR}, a synthetic retrieval training dataset containing 33 (high to very-low resource) languages for fine-tuning multilingual dense retrievers without requiring any human supervision. 
To construct SWIM-IR, we propose SAP ({\it summarize-then-ask prompting}), where the large language model (LLM) generates a textual summary prior to the query generation step. SAP assists the LLM in generating informative queries in the target language.
Using SWIM-IR, we explore synthetic fine-tuning of multilingual dense retrieval models and evaluate them robustly on three retrieval benchmarks: XOR-Retrieve (cross-lingual), MIRACL (monolingual) and XTREME-UP (cross-lingual). 
Our models, called SWIM-X, are competitive with human-supervised dense retrieval models, e.g., mContriever-X, finding that SWIM-IR can cheaply substitute for expensive human-labeled retrieval training data. SWIM-IR dataset and SWIM-X models are available at: \url{https://github.com/google-research-datasets/SWIM-IR}.
\end{abstract}

\begin{table}[]
 \centering
\resizebox{\textwidth}{!}{
 \begin{tabular}{llcccrr}
\toprule
\textbf{Dataset} &  \textbf{Q Gen.} & \textbf{Cross.} & \textbf{Mono.} & \textbf{\#~L } & \textbf{Domain} & \textbf{\#~Train} \\ \midrule
    NeuCLIR & Human & EN$ \rightarrow$L & L$ \rightarrow$L & 3 & News (hc4) & \textcolor{red}{$\boldsymbol{\times}$} \\
    MKQA & Human & L$ \rightarrow$EN & \textcolor{red}{$\boldsymbol{\times}$} & 26 & Wikipedia & 10K \\
    mMARCO  & Translate & \textcolor{red}{$\boldsymbol{\times}$} & L$ \rightarrow$L & 13 & MS MARCO & 533K \\
    Mr.TyDI & Human & \textcolor{red}{$\boldsymbol{\times}$} & L$ \rightarrow$L & 11 & Wikipedia & 49K \\
    MIRACL & Human & \textcolor{red}{$\boldsymbol{\times}$} & L$ \rightarrow$L & 18 & Wikipedia & 726K \\
    JH-POLO & GPT-3 & EN$ \rightarrow$L & \textcolor{red}{$\boldsymbol{\times}$} & 3 & News (hc4) & 78K \\
    \rowcolor{paleaqua} \textbf{SWIM-IR} & \textbf{PaLM 2} & \textbf{L$ \rightarrow$EN} & \textbf{L$ \rightarrow$L} & \textbf{33} & \textbf{Wikipedia} & \textbf{28M} \\
    \bottomrule
    \end{tabular}}
    \vspace*{-2.5mm}
    \caption{We construct SWIM-IR, a ``synthetic'' multilingual dataset with 28 million PaLM 2~generated training pairs across 33 languages in our work;~(Q Gen.)~denotes the query generation technique; (Cross.~and Mono.)~denotes the retrieval task and (query$\rightarrow$document) language pair; (\#~L and \#~Train)~denotes the language count and available training pairs.}
    \label{tab:dataset-comparison-small}
\end{table}

\begin{figure}[t]
    \centering
    \begin{center}
        \includegraphics[trim=0 5 0 0,clip,width=
        \textwidth]{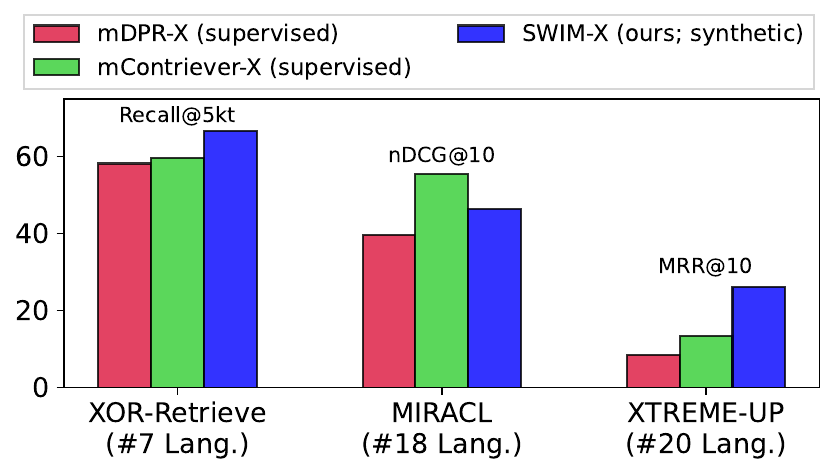}
        \caption{Summary of the quantitative results across three multilingual retrieval benchmarks evaluated in our work. SWIM-X is fine-tuned on SWIM-IR (PaLM 2 generated synthetic training data) without any human supervision. All scores are macro-averaged. \vspace*{-3mm}}
        \label{fig:overall_results}
        \vspace*{-4mm}
    \end{center}
\end{figure}

\begin{figure*}[t]
    \centering
    \begin{center}
        \includegraphics[trim=90 490 90 210,clip,width=\textwidth]{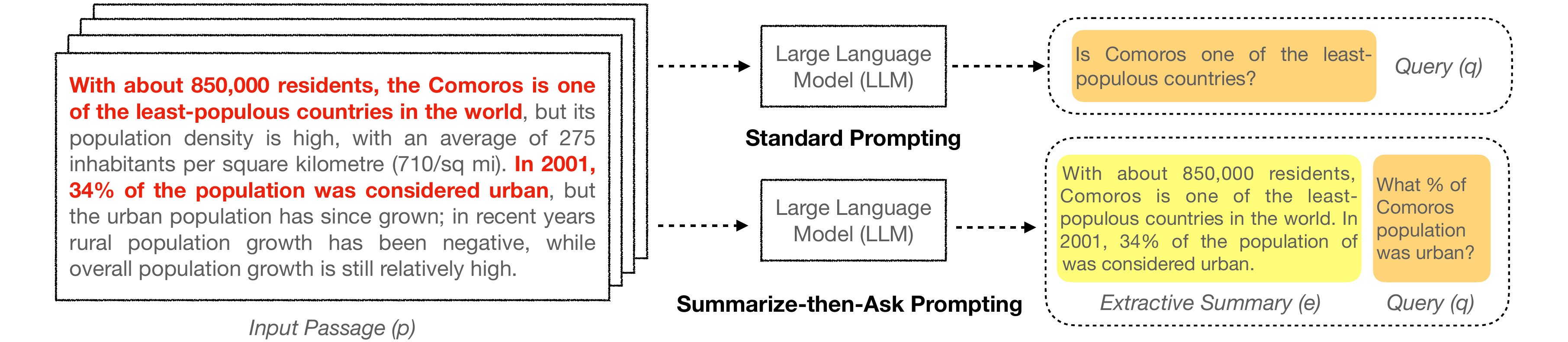}
        \caption{An illustration of SAP ({\it Summarize-then-Ask Prompting}) versus standard prompting for English query generation on English Wikipedia. SAP assists the LLM in improving the query generation quality (orange box) by identifying the relevant sections of the input passage (highlighted in red) via the extractive summarization (yellow box) as an intermediate reasoning step. \vspace{-3mm}}
        \label{fig:sap-vs-standard}
    \end{center}
    \vspace*{-\baselineskip}
\end{figure*}

\section{Introduction}

Dense retrieval models have demonstrated impressive performance in ad-hoc information retrieval (IR) tasks, e.g., web search, outperforming traditional retrieval systems such as BM25 \cite[\emph{inter alia}]{karpukhin-etal-2020-dense, lin2021pretrained, Ni2022LargeDE, neelakantan2022text}. 
A major reason for its success lies in the availability of large-scale supervised training datasets in English, such as MS MARCO \cite{msmarco} or NQ \cite{nq}, and coupled with effective training strategies, such as custom hard-negative mining \cite{xiong:2021, lin:2023}, or teacher distillation \cite{hoeffstater:2021, ren:2021}.

However, there is a limited exploration of dense retrieval models in multilingual retrieval,\footnote{Throughout the paper, we use ``{multilingual retrieval}'' to collectively denote both cross-language, i.e., cross-lingual and within language, i.e., monolingual retrieval tasks.} due to uneven and low distribution of human-supervised training data for other languages apart from English \cite{reimers:2020, feng:2022,wieting-etal-2023-beyond}.
Collecting human annotations for training data generation is not scalable, as it is cumbersome to search and hire native speakers, check their language proficiency, and teach them. Additionally, human annotators are expensive, thereby requiring a large annotation budget for generating a sufficient amount of training pairs (cf.~\autoref{fig:swim-training-data}).

Multilingual query generation is a complex task \cite{wang:2021}. It requires understanding of semantic mappings of words across languages, similar to machine translation \cite{forcada:2002, tan-etal-2019-multilingual, zhu:2023}. Recently, utilizing LLMs for query generation has been popular in English \cite{Bonifacio2022InParsUD, Dai2022PromptagatorFD}.
But as illustrated in \autoref{fig:sap-vs-standard}, standard prompt templates can lead the LLM to generate either extractive or uninformative\footnote{{\it Uninformative} denotes a query that can be easily answered using the first (or last) few words in the passage.} queries across languages.

\begin{figure*}[t]
    \centering
    \begin{center}
        \includegraphics[trim=0 0 0 0,clip,width=\textwidth]{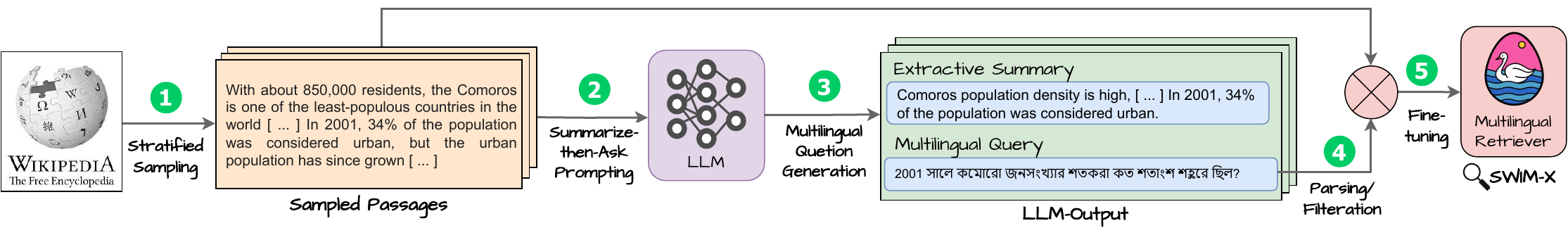}
        \caption{An illustration of the cross-lingual SWIM-IR~dataset construction procedure. Steps are as follows: (1) Sample N passages from the English Wikipedia using stratified sampling for each language out of the L languages; (2) Feed a sampled passage along with the few-shot exemplars to the LLM with SAP; (3 \& 4) Parse the LLM output to receive the synthetic query in the target language (above in Bengali); (5) Fine-tune a multilingual dense retriever model (SWIM-X) with training pairs combined for all languages, i.e., N$\times$L pairs. \vspace{-2mm}}
        \label{fig:SWIM-IR-diagram}
    \end{center}
    \vspace*{-\baselineskip}
\end{figure*}

To improve the quality of the generated query, we propose SAP ({\it Summarize-then-Ask Prompting}), where we optimize the prompt to break down the query generation with LLM in two stages: (i) {\it summary extraction}, which identifies the relevant information from the long input passage and extracts the best representative sentences as the summary, and (ii) {\it query generation}, which generates a multilingual query relevant for the input passage, using the extracted summary (first stage) as the intermediate step. SAP highlights the relevant information within the passage and produces difficult (i.e., informative) queries in the target language.

In our work, we utilize PaLM 2 \cite{anil2023palm}, a recent multilingual LLM (successor of PaLM 540B \cite{chowdhery2022palm}) for query generation. The generated query paired with the original passage from Wikipedia is used to construct the SWIM-IR dataset.
SWIM-IR provides synthetic training (query-passage) pairs for improving dense retrieval models without requiring any human supervision. The dataset spans across 33 diverse languages, including both high and very-low resource languages and is one of the largest multilingual synthetic training dataset with 28 million training pairs (cf. \autoref{tab:dataset-comparison-small}).

We develop synthetic multilingual (both monolingual and cross-lingual) dense retrieval models called SWIM-X, using mT5 (base) \cite{xue2021mt5} as the backbone and fine-tune on SWIM-IR. We compare SWIM-X against models fine-tuned with human supervision by changing only the training dataset while keeping other, i.e., model parameters and training settings unchanged. 
We evaluate on three standard multilingual retrieval benchmarks (two cross-lingual and one monolingual).
As shown in \autoref{fig:overall_results}, on XOR-Retrieve \cite{asai-etal-2021-xor}, SWIM-X outperforms the best-supervised baseline (mContriever-X) by 7.1 points at Recall@5kt. On MIRACL \cite{miracl}, a monolingual retrieval benchmark, SWIM-X is inferior to mContriever-X by 9.0 points at nDCG@10, which shows room for future improvement. On XTREME-UP \cite{ruder2023xtremeup}, a challenging benchmark containing 20 underrepresented Indo-European languages, SWIM-X~outperforms mContriever-X by 11.7 points at MRR@10. We publicly open-source SWIM-IR dataset and SWIM-X models at \url{https://github.com/google-research-datasets/SWIM-IR}.

\vspace{-1mm}
\section{SWIM-IR Dataset Overview}
\vspace{-1mm}
In our dataset overview, we first describe the SAP design formulation for multilingual query generation (\S\ref{subsec:prompting-design}), data construction details (\S\ref{subsec:data-construction}), and finally discuss about human validation and content filtration  (\S\ref{subsec:dataset-analysis}).

\vspace{-1mm}
\subsection{SAP Design Formulation}\label{subsec:prompting-design}
Multilingual query generation is not a trivial task as it requires a deep understanding of the passage content and its own translations across different languages \cite{wang:2021}. Also, passages can often be lengthy and contain information on multiple topics. Using the entire passage can potentially cause hallucinations in models by generating non-meaningful queries, which affects the retrieval performance \cite{gospodinov:2023}.

To break down the task complexity of multilingual query generation and improve the query quality, we implement summarize-then-ask prompting (SAP). As shown above in \autoref{fig:sap-vs-standard}, we identify the relevant information within a passage by asking the LLM to generate an extractive summary and use it as an intermediate step for generating informative queries \cite{wei2023chainofthought}. The procedure is described in more detail below: 

\smallskip
\noindent\textbf{(i) Summary extraction.} The LLM constructs an extractive summary $e_s$ of the input passage $p_{s}$, where $s$ denotes the source language. 
The summary captures the highly relevant information contained within the passage $p_{s}$ (which occasionally may be long) acting as an useful intermediate signal for the LLM to generate a multilingual query in the later stage. We denote the first stage as $e_s = \mathrm{LLM}(p_s; \theta^1, \cdots, \theta^k)$, where ($\theta^1, \cdots, \theta^k$) denotes the $k$ few-shot prompt exemplars\footnote{Multilingual query generation requires few-shot prompt exemplars. As our experiments show in (\S\ref{sec:ablation_studies}), zero-shot prompting often generates unparseable outputs with PaLM 2.} containing the passage, summary in the source language $s$ and the query in the target language $t$.\footnote{In our work, we did not use abstractive summarization, as LLMs have notoriously been shown to hallucinate and generate factual inconsistencies in their output generations \cite{maynez-etal-2020-faithfulness, liu2023evaluating}.}

\smallskip
\noindent\textbf{(ii) Query generation.} Next, the LLM combines the summary $e_{s}$ generated previously with the original input passage $p_{s}$, highlighting the relevant information required for composing the query $q_{t}$ in the target language $t$. We denote this stage as $q_t = \mathrm{LLM}(e_{s},p_{s}; \theta^1, \cdots, \theta^k)$, where extractive summary $e_s$, input passage $p_{s}$ and $k$-shot exemplars all appear from the first stage.

\subsection{SWIM-IR Dataset Construction}\label{subsec:data-construction}
For constructing SWIM-IR, we only require an unlabeled corpus of passages and few-shot exemplars. An overview of the cross-lingual generation procedure is shown in \autoref{fig:SWIM-IR-diagram}. Prompt examples are provided in the Appendix (\S\ref{sec:prompts}).

\begin{table*}[t]
\centering
\resizebox{\textwidth}{!}{
\begin{tabular}{l c c c c c c c c c c}
        \toprule
         \textbf{Benchmark} & \textbf{Retrieval}& \textbf{Evaluation} & \textbf{Query $\rightarrow$ Passage} & \textbf{\#~L} & \textbf{ISO}
        & \multicolumn{1}{c}{\textbf{Languages}} & \multicolumn{2}{c}{\textbf{Train Split}} & \multicolumn{2}{c}{\textbf{Dev/Test Split}} \\
         & \textbf{Task} & \textbf{Metric} & & & & & \textbf{{\#~Q}} & \textbf{HNeg.} & \textbf{{\#~Q}} & \textbf{\#~Passages} \\ \midrule
        \multicolumn{1}{p{3cm}}{XOR-Retrieve \cite{asai-etal-2021-xor}} & Cross-lingual & Recall@5kt & L $\rightarrow$ English & 7 & \multicolumn{1}{p{3cm}}{\texttt{ar}, \texttt{bn}, \texttt{fi}, \texttt{ja}, \texttt{ko}, \texttt{ru}, \texttt{te}} & \multicolumn{1}{p{7cm}}{Arabic, Bengali, Finnish, Japanese, Korean, Russian, Telugu} & 15,250 & Yes (1 each) & 2,110 & 18,003,200 \\ \midrule
        \multicolumn{1}{p{3cm}}{MIRACL \cite{miracl}} & Monolingual & nDCG@10 & L $\rightarrow$ L & 18 & \multicolumn{1}{p{3cm}}{\texttt{ar}, \texttt{bn},  \texttt{de}, \texttt{en}, \texttt{es}, \texttt{fa}, \texttt{fi}, \texttt{fr}, \texttt{hi}, \texttt{id}, \texttt{ja}, \texttt{ko}, \texttt{ru}, \texttt{sw}, \texttt{te}, \texttt{th}, \texttt{yo}, \texttt{zh}}
        & \multicolumn{1}{p{7cm}}{Arabic, Bengali, German, English, Spanish, Farsi, Finnish, French, Hindi, Indonesian, Japanese, Korean, Russian, Swahili, Telugu, Thai, Yoruba, Chinese} & 88,288 & Yes (max 4) & 13,495 & 106,332,152 \\ \midrule
        \multicolumn{1}{p{3cm}}{XTREME-UP \cite{ruder2023xtremeup}} & Cross-lingual & MRR@10 & L $\rightarrow$ English & 20 & \multicolumn{1}{p{3cm}}{\texttt{as}, \texttt{bho}, \texttt{brx}, \texttt{gbm}, \texttt{gom}, \texttt{gu}, \texttt{hi}, \texttt{hne}, \texttt{kn}, \texttt{mai}, \texttt{ml}, \texttt{mni}, \texttt{mr}, \texttt{mwr}, \texttt{or}, \texttt{pa}, \texttt{ps}, \texttt{sa}, \texttt{ta}, \texttt{ur}} & \multicolumn{1}{p{7cm}}{Assamese, Bhojpuri, Boro, Garhwali, Konkani, Gujarati, Hindi, Chhattisgarhi, Kannada, Maithili, Malayalam, Manipuri, Marathi, Marwari, Odia, Punjabi, Pashto, Sanskrit, Tamil, Urdu} & 13,270 & No & 5,300 &  112,426 \\
        \bottomrule
    \end{tabular}}
    \vspace{1mm}
    \caption{Overview of the multilingual retrieval evaluation benchmarks used in our work: (i) XOR-Retrieve (Dev) \cite{asai-etal-2021-xor}, (ii) MIRACL (Dev) \cite{miracl} and (iii) XTREME-UP (Test) \cite{ruder2023xtremeup}; (HNeg.)~denotes availability of hard negatives for fine-tuning; (\#~L) denotes the number of languages covered by the benchmark; (\#~Q)~denotes the number of queries in each dataset split.}
    \vspace*{-0.75\baselineskip}
    \label{tab:evaluation-datasets}
\end{table*}
\begin{table}[t]
\centering
\resizebox{\textwidth}{!}{
\begin{tabular}{l@{\hskip 0.2in} c c c@{\hskip 0.2in} c c c@{\hskip 0.2in} c c c }
        \toprule
        \textbf{Lang. (ISO)} & \multicolumn{3}{c}{\textbf{fluency ($\uparrow$)}} & \multicolumn{3}{c}{\textbf{adequacy ($\uparrow$)}} & \multicolumn{3}{c}{\textbf{language ($\uparrow$)}} \\
        \cmidrule(lr{1.0em}){2-4} \cmidrule(lr{1.0em}){5-7} \cmidrule(lr{1.0em}){8-10}
        \multicolumn{1}{l}{\textbf{Rating ($\rightarrow$)}} & \textbf{0} & \textbf{1} & \textbf{2} & \textbf{0} & \textbf{1} & \textbf{2} & \textbf{0} & \textbf{1} & \textbf{2} \\
        \midrule
        English (\texttt{en}) & 2\% & 3\% & \green{95\%}  & 2\% & \red{13\%} & \green{85\%} &  0\% & 0\% & \green{100\%} \\ 
        Spanish (\texttt{es}) & 1\% & \red{10\%} & \green{89\%} & \red{14\%} & \red{12\%} & 74\% & 1\% & 0\% & \green{99\%} \\
        Chinese (\texttt{zh}) & 7\% & \red{19\%} & 74\% & 7\% & \red{30\%} & 63\% & 0\% & 0\% & \green{100\%} \\
        Hindi (\texttt{hi})   & \red{12\%} & 5\% & \green{83\%} & 6\% & \red{19\%} & \green{75\%} & 0\% & 0\% & \green{100\%}  \\
        Bengali (\texttt{bn}) & 6\% & 4\% & \green{90\%} & \red{10\%} & \red{14\%} & \green{76\%} & 1\% & 0\% & \green{99\%} \\
        \bottomrule
    \end{tabular}}
    \vspace*{-2.5mm}
    \caption{Human validation statistics on SWIM-IR. Annotators evaluate the quality of the generated query on a three-level rating scale (0/1/2) based on thee factors: (i) fluency, (ii) adequacy and (iii) language.
    \vspace{-0.75\baselineskip}
    \label{tab:swim-ir-question-quality}
    }
\end{table}

\smallskip
\noindent\textbf{Cross-lingual.} The goal is to generate a query in the target language $t$ using the input passage in English (source language $s$). 
We use a stratified sampling algorithm (for more details, refer to \S\ref{sec:sampling-strategy} in the Appendix) to
sample a maximum of one million passages for each target language $t$ from the English Wikipedia corpus used in XOR-Retrieve \cite{clark-etal-2020-tydi, asai-etal-2021-xor} or XTREME-UP \cite{ruder2023xtremeup}.
Next, we construct five prompt exemplars and manually construct both the summary and query for the exemplar in English. Further, we use Google Translate\footnote{Google Translate: \href{https://translate.google.com/}{translate.google.com}} to translate the exemplar queries across other languages. Finally, we construct the prompt, where we explain our query generation task as an instruction, include the target language, and the 5-shot exemplars as an input to the LLM with SAP.

\smallskip
\noindent\textbf{Monolingual.} The goal is to generate a query in the same language as the input passage ($s=t$). 
We follow the setting similar to the cross-lingual task. We first sample one million passages (if available) for each language-specific Wikipedia corpus in MIRACL \cite{miracl}.\footnote{For 16 out of the 18 languages, MIRACL contains a training split except for two: German (\texttt{de}) and Yoruba (\texttt{yo}).} Next, we carefully select three training pairs as our prompt exemplars.\footnote{As language-specific passages consume more tokens, e.g., Telugu, to save computational budget, we rely only on 3-shot exemplars (instead of five) for the monolingual task.} For languages with no training split, we manually construct our prompt exemplars. Further, we use Google Bard\footnote{Google Bard: \href{https://bard.google.com/}{bard.google.com}} to generate exemplar summaries in the target language. Finally, we construct the prompt, where we explain our query generation task as an instruction, and the 5-shot exemplars with SAP. 

\begin{table*}[t]
    \centering
    \small    
    \resizebox{0.95\columnwidth}{!}{%
    \setlength\tabcolsep{3.5pt}
    \begin{tabular}{l c c c c|ccccccc}
       \toprule
       Model & PLM & PT & Finetune & \multicolumn{8}{c}{Recall@5kt} \\
       & & & (Datasets) & Avg. & Ar & Bn & Fi & Ja & Ko & Ru & Te \\ 
      \specialrule{.4pt}{2pt}{0pt}
       \rowcolor{paleaqua} \multicolumn{12}{l}{\textit{Existing Supervised Baselines (Prior work)}} \\ 
    %   \specialrule{.4pt}{0pt}{2pt}
       Dr. DECR \cite{dr-decr} & XLM-R & WikiM & \scriptsize{NQ + XOR$^*$}    & 73.1 & 70.2 & 85.9 & 69.4 & 65.1 & 68.8 & 68.8 & 83.2 \\ 
       mDPR \cite{asai-etal-2021-xor} & mBERT & --- & \scriptsize{XOR}  & 50.2 & 48.9 & 60.2 & 59.2 & 34.9 & 49.8 & 43.0 & 55.5 \\ 
       mBERT + xQG \cite{Zhuang2023AugmentingPR} & mBERT & --- & \scriptsize{XOR} & 53.5 & 42.4 & 54.9 & 54.1 & 33.6 & 52.3 & 33.8 & 52.5 \\ \specialrule{.4pt}{0pt}{2pt}
       Google MT + DPR \cite{asai-etal-2021-xor} & BERT & --- & \scriptsize{NQ} & 69.6 & 69.6 & 82.2 & 62.4 & 64.7 & 68.8 & 60.8 & 79.0 \\
       OPUS MT + DPR \cite{asai-etal-2021-xor} & BERT & --- & \scriptsize{NQ} & 50.6 & 52.4 & 62.8 & 61.8 & 48.1 & 58.6 & 37.8 & 32.4  \\ 
    %   \specialrule{.4pt}{2pt}{0pt}
       \rowcolor{paleaqua} \multicolumn{12}{l}{\textit{Zero-shot baselines (English-only supervision)}} \\ 
    %   \specialrule{.4pt}{0pt}{2pt}
       mContriever & mT5 & mC4 & --- & 38.9 & 35.9 & 33.9 & 43.6 & 34 & 35.1 & 45.1 & 44.5 \\
       mDPR-EN  & mT5 & --- & \scriptsize{MS MARCO} & 39.3 & 34.3 & 35.5 & 45.2 & 40.2 & 36.5 & 43.9 & 39.5 \\ 
       mContriever-EN & mT5 & mC4 & \scriptsize{MS MARCO} & 44.0 & 37.5 & 38.2 & 50.6 & 41.1 & 37.2 & 49.8 & 53.8 \\       
    %   \specialrule{.4pt}{2pt}{0pt}
       \rowcolor{paleaqua} \multicolumn{12}{l}{\textit{Supervised Baselines (Cross-lingual supervision)}} \\ 
    %   \specialrule{.4pt}{0pt}{2pt}
       mDPR-X  & mT5 & --- & \scriptsize{XOR} & 53.6 & 51.5 & 63.5 & 52.5 & 45.6 & 52.3 & 43.0 & 66.8 \\
       mContriever-X  & mT5 & mC4 & \scriptsize{XOR} & 55.3 & 52.1 & 68.1 & 54.5 & 47.7 & 50.5 & 50.2 & 64.3 \\
       \specialrule{.4pt}{0pt}{2pt}
       mDPR-X & mT5 & --- & \scriptsize{MS MARCO + XOR} & 58.2 & 55.3 & 70.1 & 56.7 & 49.8 & 55.8 & 50.6 & 69.3 \\ 
       mContriever-X & mT5 & mC4 & \scriptsize{MS MARCO + XOR} & 59.6 & 54.7 & 73.4 & 57.0 & 53.1 & 56.5 & 51.5 & 71.0 \\ 
    %   \specialrule{.4pt}{2pt}{0pt}
       \rowcolor{paleaqua} \multicolumn{12}{l}{\textit{Synthetic Baselines (Our work)}} \\ 
    %   \specialrule{.4pt}{0pt}{2pt}
       SWIM-X (500K)  & mT5 & --- & \scriptsize{SWIM-IR} & 59.0 & 54.0 & 67.4 & 59.2 & 52.7 & 55.1 & 54.4 & 70.2 \\
       SWIM-X (500K)  & mT5 & mC4 & \scriptsize{SWIM-IR} & 63.0 & 57.0 & 71.1 & 61.8 & 56.8 & 60.7 & 63.3 & 70.2   \\
       \specialrule{.4pt}{0pt}{2pt}
       SWIM-X (7M)  & mT5 & --- & \scriptsize{SWIM-IR} & 65.1 & 57.9 & 75.0 & 65.6 & 59.3 & 58.9 & 64.6 & 74.4 \\ 
       SWIM-X (7M)  & mT5 & mC4 & \scriptsize{SWIM-IR} & 66.7 & 61.2 & 77.0 & 65.0 & 62.2 & 62.8 & 65.4 & 73.5 \\
       \bottomrule
    \end{tabular}}
    \caption{Experimental results showing Recall@5kt for cross-lingual retrieval on XOR-Retrieve dev \cite{asai-etal-2021-xor}; (PLM) denotes the pre-trained language model; (PT) denotes the pre-training dataset; ($^*$) Dr.DECR is fine-tuned in a complex training setup across more datasets ($\mathsection$\ref{sec:implementation-details}); WikiM denotes WikiMatrix \cite{schwenk-etal-2021-wikimatrix}; XOR denotes XOR-Retrieve; SWIM-X (ours) is fine-tuned on 500K and 7M synthetic data. \vspace{-2mm}}
    \label{tab:xor-retrieve-results}
    \vspace{-2mm}
\end{table*}

\subsection{Human Validation \& Content Filtration}\label{subsec:dataset-analysis}

\noindent\textbf{Human validation.} The goal of our query generation is to generate an adequate and fluent query according to a given passage \cite{qiu-xiong-2019-generating}. To evaluate the intrinsic query quality, we conduct a validation study in SWIM-IR on a subset of five languages.\footnote{The authors in the paper are native speakers of the five languages used for evaluation: English (\texttt{en}), Bengali (\texttt{bn}), Spanish (\texttt{es}), Chinese (\texttt{zh}) and Hindi (\texttt{hi}).}
Within the five evaluated languages, three are high-resource, one medium-resource and one low-resource. For each language, we randomly sample a fixed amount of query-passage pairs resulting in a overall sum of 500 evaluation pairs to be human validated across all languages.

We compute the query quality on a three-level rating scheme (0/1/2) based on three evaluation criteria: fluency, adequacy, and language. (i) \emph{fluency}, measures the coherence of the generated query, i.e., whether the query is understandable and readable by the user and contains no spelling or grammatical mistakes. (ii) \emph{adequacy}, measures the relevancy of the query with passage (used for query generation) (iii) \textit{language}, detects the language of the generated query, or whether code-switching occurs in the generated query.

\smallskip
\noindent\textbf{Validation statistics.} \autoref{tab:swim-ir-question-quality} reports the human validation statistics. For fluency, major mistakes are observed in Hindi (12\%) and Chinese (7\%), where the passage sampled in MIRACL \cite{miracl} can be too short (only 2--3 words long), this leads to the exact duplication of the exact text in the query. For adequacy, we observe that in Chinese (30\%) of the queries are not relevant to the passage. Similar to fluency, a low adequacy is observed in cases when either query is generated for a short passage or when the query is about a related topic which is not directly referenced within the passage. Finally for language, annotators achieve between 99--100\% for all languages indicating PaLM 2 is likely to generate the query in the correct language.

\smallskip
\noindent\textbf{Content filtration}. LLMs have been shown to generate undesirable content, particularly under conditions that prime the model with material targeted at drawing out any negative patterns or associations in the training data \cite{gehman:2020, bender:2021}. To avoid this, we 
use the Google Cloud Natural Language content classification categories\footnote{\href{https://cloud.google.com/natural-language/docs/categories}{cloud.google.com/natural-language/docs/categories}} to 
filter out harmful content present within the SWIM-IR training pairs. We discard samples with a high content classification of either \texttt{/Adult} or any of the \texttt{/Sensitive~Subjects} labels. For more details on content filtration, refer to (\S\ref{sec:content-filtering}) in the Appendix.

\vspace{-2mm}
\section{Experiments}

\subsection{Datasets and Metrics} We evaluate on three multilingual retrieval benchmarks: (i) \textbf{XOR-Retrieve} \cite{asai-etal-2021-xor}, (ii) \textbf{MIRACL} \cite{miracl} and (iii) \textbf{XTREME-UP} \cite{ruder2023xtremeup}. XOR-Retrieve and XTREME-UP are cross-lingual and MIRACL is monolingual. Following prior work, we evaluate models at Recall@5kt on XOR-Retrieve, nDCG@10 on MIRACL and MRR@10 on XTREME-UP. An overview of the evaluation dataset statistics is available in \autoref{tab:evaluation-datasets}.
For additional details, refer to the Appendix (\S\ref{sec:SWIM-IR-evaluation}). 

\begin{table*}[]
\centering
\resizebox{\textwidth}{!}{
\begin{tabular}{l|c|cccccccccccccccccc}
\toprule
% \multicolumn{2}{c}{\textbf{MIRACL}} & \multicolumn{4}{c}{\textbf{Existing Baselines}} &  \multicolumn{2}{c}{\textbf{Zero-shot (En)}} & \multicolumn{2}{c}{\textbf{Gold FT}} & \multicolumn{1}{c}{\textbf{Synthetic FT}} \\
% Model & & \multicolumn{18}{c}{\textbf{nDCG@10} on MIRACL \cite{miracl}} \\
 Model & Avg. & \texttt{ar} & \texttt{bn} & \texttt{en} & \texttt{es} & \texttt{fa} & \texttt{fi} & \texttt{fr} & \texttt{hi} & \texttt{id} & \texttt{ja} & \texttt{ko} & \texttt{ru} & \texttt{sw} & \texttt{te} & \texttt{th} & \texttt{zh} & \texttt{de} & \texttt{yo} \\
\specialrule{.4pt}{2pt}{0pt}
\rowcolor{paleaqua} \multicolumn{20}{l}{\textit{Existing Supervised Baselines (Prior work)}} \\ 
% \specialrule{.4pt}{0pt}{2pt}
BM25 & 38.5 & 48.1 & 50.8 & 35.1 & 31.9 & 33.3 & 55.1 & 18.3 & 45.8 & 44.9 & 36.9 & 41.9 & 33.4 & 38.3 & 49.4 & 48.4 & 18.0 & 22.6 & 40.6 \\
mDPR & 41.8 & 49.9 & 44.3 & 39.4 & 47.8 & 48.0 & 47.2 & 43.5 & 38.3 & 27.2 & 43.9 & 41.9 & 40.7 & 29.9 & 35.6 & 35.8 & 51.2 & 49.0 & 39.6 \\
Hybrid & 56.6 & 67.3 & 65.4 & 54.9 & 64.1 & 59.4 & 67.2 & 52.3 & 61.6 & 44.3 & 57.6 & 60.9 & 53.2 & 44.6 & 60.2 & 59.9 & 52.6 & 56.5 & 37.4 \\
Cohere-API & 54.2 & 66.7 & 63.4 & 50.1 & 50.7 & 48.4 & 67.5 & 44.3 & 57.3 & 50.5 & 51.6 & 54.6 & 47.7 & 54.3 & 63.8 & 60.6 & 38.9 & 41.4 & 62.9 \\
% \specialrule{.4pt}{2pt}{0pt}
\rowcolor{paleaqua} \multicolumn{20}{l}{\textit{Zero-shot baselines (English-only supervision)}} \\
% \specialrule{.4pt}{0pt}{2pt}
mDPR-EN & 39.8 & 49.7 & 50.1 & 35.4 & 35.3 & 39.3 & 48.2 & 31.3 & 37.4 & 35.6 & 38.9 & 44.1 & 36.1 & 33.8 & 49.2 & 50.6 & 34.7 & 32.1 & 34.4 \\
mContriever-EN & 37.8 & 49.1 & 48.4 & 32.7 & 33.3 & 37.1 & 48.4 & 27.0 & 35.9 & 32.7 & 34.1 & 40.2 & 35.1 & 44.5 & 46.2 & 45.0 & 27.5 & 29.7 & 33.7 \\
% \specialrule{.4pt}{2pt}{0pt}
\rowcolor{paleaqua} \multicolumn{20}{l}{\textit{Supervised Baselines (Monolingual supervision)}} \\
% \specialrule{.4pt}{0pt}{2pt}
mDPR-X & 39.6 & 52.8 & 57.1 & 30.2 & 24.7 & 37.6 & 46.1 & 26.4 & 27.8 & 37.3 & 42.9 & 38.3 & 34.9 & 53.7 & 68.4 & 58.2 & 34.9 & 19.2 & 22.2 \\
mContriever-X & 55.4 & 66.4 & 68.4 & 44.2 & 42.8 & 48.9 & 65.2 & 46.2 & 45.0 & 45.8 & 56.8 & 58.8 & 51.2 & 67.7 & 79.0 & 70.7 & 49.4 & 42.3 & 48.4 \\
% \specialrule{.4pt}{2pt}{0pt}
\rowcolor{paleaqua} \multicolumn{20}{l}{\textit{Synthetic Baselines (Our work)}} \\
% \specialrule{.4pt}{0pt}{2pt}
SWIM-X (180K) & 46.4 & 60.2 & 57.1 & 34.7 & 33.4 & 36.3 & 40.6 & 64.3 & 33.0 & 39.5 & 40.8 & 43.3 & 49.7 & 40.0 & 55.9 & 56.3 & 63.3 & 50.2 & 36.5 \\
\bottomrule
\end{tabular}
}
\caption{
Experimental results for monolingual retrieval on MIRACL dev \cite{miracl}. All scores denote \textbf{nDCG@10}; (Hyb.)~denotes Hybrid retriever with ranked fusion of three retrievers: mDPR, mColBERT and BM25; BM25, mDPR and Hybrid scores taken from \cite{miracl}; Cohere-API is used as a reranker on top of 100 BM25 results, taken from \cite{kamalloo2023evaluating}. SWIM-X (ours) is fine-tuned on 180K synthetic training pairs. \vspace{-2mm}}
\label{tab:miracl-results-inverted}
\vspace{-2mm}
\end{table*}

\subsection{Experimental Methods} 

\textbf{Baseline categories.} Following common practice across all datasets, we evaluate three range of baselines: (i) \textit{Zero-shot baselines}: where the model denoted by ``EN'' (model-EN) is fine-tuned using supervised English-only training data such as MS MARCO \cite{msmarco} or NQ \cite{nq}. (ii) \textit{Supervised baselines}: where the model denoted by ``X'' (model-X) is fine-tuned on human-supervised, i.e.,~multilingual training data. (iii) \textit{Synthetic baselines}: where the model denoted by ``SWIM-X'' is fine-tuned without any supervision, relying purely on synthetic multilingual training data. Additionally, we report the amount of synthetic pairs, e.g., SWIM-X (500K) is fine-tuned on 500K training pairs.

\smallskip
\noindent\textbf{Model choices.} For our dense retrieval models, we adapt DPR \cite{karpukhin-etal-2020-dense} to the multilingual setting with the mT5-base \cite{xue2021mt5} language model with 580M parameters. 
Next, we include mContriever \cite{izacard2022unsupervised} which adopts an additional pre-training stage with contrastive loss based on unsupervised data prepared from pairwise sentence cropping in mC4~\cite{xue2021mt5}. For query generation, we use PaLM 2 (S) \cite{anil2023palm} for efficient generation due to its low-cost and inference latency.

\smallskip
\noindent\textbf{Existing baselines.} For XOR-Retrieve, we include Dr.~DECR \cite{dr-decr}, a cross-lingual ColBERT \cite{colbert} fine-tuned on a large amount of supervised data in a computationally expensive setup involving knowledge distillation with English ColBERTv2 \cite{santhanam-etal-2022-colbertv2}.~xQG \cite{Zhuang2023AugmentingPR} involves cross-language query generation and concatenating the queries along with the passage representation. We also include two-stage translation baselines, Google Translate and Opus-MT from \citet{asai-etal-2021-xor}. For MIRACL, we include the official BM25, mDPR and Hybrid (combining BM25, mDPR and mColBERT) baselines \cite{miracl}, and Cohere-API is used as a reranker with top-100 BM25 results \cite{kamalloo2023evaluating}.

\vspace{-2mm}
\subsection{Training Methodology}\label{sec:implementation-details} 
\vspace{-1mm}

\textbf{Zero-shot \& supervised baselines.} We replicate mContriever and mDPR zero-shot baselines by initializing from an mT5-base checkpoint \cite{xue2021mt5} and further fine-tuning on MS MARCO, following a setup similar to \citet{Ni2022LargeDE}. Similarly, mContriever-X and mDPR-X have been additionally fine-tuned on training split available for each dataset. For additional technical details on supervised baselines, refer to the Appendix (\S\ref{sec:baseline-models}). As mContriever includes an additional pre-training stage, we set the batch size to 8192, 
learning rate to $1e^{-3}$ and pre-train for 600K steps with mC4 \cite{xue2021mt5}. For more details on pre-training, refer to the Appendix (\S\ref{sec:mc4-pretraining}).

\smallskip
\noindent\textbf{Synthetic baselines.} For SWIM-X, we pre-train the mT5-base checkpoint on mC4 \cite{xue2021mt5} for 600K steps using a contrastive loss function objective, similar to Contriever \cite{izacard2022unsupervised}.  Next, we fine-tune the pre-trained mT5-base model on SWIM-IR with in-batch negatives and a contrastive loss function \cite{oord:2018}. During fine-tuning, we set the batch size to 4096, learning rate to $1e^{-3}$ and fine-tune between 5K to 50K training steps, depending upon the size of the training dataset. For technical details on synthetic baselines, refer to the Appendix (\S\ref{sec:synthetic-baselines}).

\subsection{Experimental Results}

\textbf{XOR-Retrieve.} \autoref{tab:xor-retrieve-results} shows that SWIM-X (7M), fine-tuned on 7M synthetic pairs (max.~of 1M per language) outperforms the best supervised baseline, mContriever-X, by 7.1 points Recall@5kt. Without mC4 pre-training, SWIM-X (7M) performance drops by only 1.6 points. We also evaluate SWIM-X (500K), a limited-budget baseline fine-tuned on 500K training pairs, which outperforms mContriever-X by 3.6 points. Few existing baselines outperform SWIM-X, however, the comparison is not fair. For instance, Dr. DECR is a multilingual ColBERT \cite{colbert} model, which  is computationally expensive at inference \cite{thakur2021beir}. Similarly, Google MT + DPR relies on a Google Translate system for the translation of queries to English.

\smallskip
\noindent\textbf{MIRACL.} \autoref{tab:miracl-results-inverted} shows that the SWIM-X (180K) model is competitive on MIRACL. SWIM-X (180K) outperforms the best zero-shot model by 6.6 points nDCG@10. However, SWIM-X underperforms mContriever-X on MIRACL, fine-tuned on 90K human-labeled training pairs with up to four hard negatives available in MIRACL by 9.0 points nDCG@10. This highlights the difficulty in the monolingual retrieval task, as models need to rely on human-supervision for improvement. Few existing baselines outperform SWIM-X, however the comparison is not fair. The Hybrid baseline relies on information based on aggregation of three models, and for Cohere-API, the underlying model information is unknown. 

\begin{table*}[]
\centering
\small    
\resizebox{\columnwidth}{!}{%
\setlength\tabcolsep{3.5pt}
\begin{tabular}{l|c|cccccccccccccccccccc}
\toprule
Model & Avg. & \texttt{as} & \texttt{bho}& \texttt{brx}& \texttt{gbm}& \texttt{gom}& \texttt{gu} & \texttt{hi} & \texttt{hne}& \texttt{kn} & \texttt{mai}& \texttt{ml} & \texttt{mni}& \texttt{mr} & \texttt{mwr}& \texttt{or} & \texttt{pa} & \texttt{ps} & \texttt{sa} & \texttt{ta} & \texttt{ur} \\
\specialrule{.4pt}{2pt}{0pt}
\rowcolor{paleaqua} \multicolumn{22}{l}{\textit{Zero-shot baselines (English-only supervision)}} \\ 
% \specialrule{.4pt}{0pt}{2pt}
\footnotesize{mDPR-EN} & 6.3 & 2.6 & 6.4 & 0.4 & 7.2 & 1.3 & 8.6 & 13.3 & 5.2 & 10.4 & 6.4 & 12.3 & 0.2 & 8.9 & 5.8 & 0.4 & 6.0 & 5.6 & 5.2 & 10.2 & 10.0 \\
\footnotesize{mContriever-EN} & 7.9 & 7.9 & 3.2 & 7.8 & 0.3 & 9.7 & 2.2 & 11.1 & 15.2 & 8.2 & 10.6 & 8.6 & 15.6 & 0.4 & 10.7 & 8.5 & 1.1 & 10.3 & 3.3 & 5.7 & 12.9 \\
% \specialrule{.4pt}{2pt}{0pt}
\rowcolor{paleaqua} \multicolumn{22}{l}{\textit{Supervised Baselines (Cross-lingual supervision)}} \\ 
% \specialrule{.4pt}{0pt}{2pt}
\footnotesize{mDPR-X} & 8.4 & 6.7 & 9.9 & 4.8 & 10.0 & 8.7 & 8.8 & 9.1 & 9.4 & 9.0 & 10.0 & 10.5 & 4.8 & 7.8 & 9.6 & 6.9 & 8.6 & 7.4 & 8.5 & 8.1 & 9.1 \\ 
\footnotesize{mContriever-X} & 12.4 & 9.8 & 15.7 & 6.7 & 14.0 & 11.7 & 13.3 & 15.5 & 13.9 & 13.6 & 13.9 & 16.9 & 6.5 & 12.0 & 13.8 & 7.5 & 13.4 & 9.8 & 12.4 & 13.0 & 14.1 \\ 
\footnotesize{mContriever-X}$^\heartsuit$ & 13.5 & 11.6 & 15.4 & 8.0 & 16.9 & 12.3 & 15.2 & 16.7 & 15.7 & 14.7 & 15.6 & 17.4 & 7.0 & 14.2 & 14.7 & 9.1 & 13.2 & 10.1 & 14.8 & 12.1 & 14.9 \\
\rowcolor{paleaqua} \multicolumn{22}{l}{\textit{Synthetic Baselines (Our work)}} \\
\footnotesize{SWIM-X (120K)}$^{MT}$ & 26.1 & 25.2 & 29.5 & 2.1 & 30.8 & 22.1 & 31.5 & 35.8 & 31.5 & 28.7 & 32.2 & 34.6 & 2.2 & 32.7 & 27.7 & 14.8 & 30.7 & 21.0 & 28.2 & 30.6 & 29.2 \\
\footnotesize{SWIM-X (120K)} & 25.2 & 24.4 & 27.7 & 4.3 & 28.3 & 25.4 & 29.4 & 32.4 & 28.8 & 30.1 & 31.8 & 34.4 & 5.1 & 30.7 & 25.7 & 15.8 & 29.6 & 20.6 & 26.1 & 27.9 & 26.1 \\
\bottomrule
\end{tabular}}
\caption{
Experimental results for cross-lingual retrieval on XTREME-UP test \cite{ruder2023xtremeup}. ($^\heartsuit$) denotes the mContriever-X model fine-tuned without MS MARCO \cite{msmarco}; Two variants of SWIM-X considered, both fine-tuned on 120K synthetic data: (1) SWIM-X (120K)$^{MT}$ fine-tuned using Google Translate, i.e., translated prompt exemplars for 15 languages, whereas (2) SWIM-X (120K) is fine-tuned using prompt exemplars sampled from XTREME-UP training split for all languages. \vspace{-2mm}}
\label{tab:xtreme-up-results-inverted}
\vspace{-2mm}
\end{table*}

\smallskip
\noindent\textbf{XTREME-UP.} \autoref{tab:xtreme-up-results-inverted} shows the results on XTREME-UP. SWIM-X (120K) is fine-tuned by randomly selecting 5 exemplars from the XTREME-UP training dataset (human-labeled queries) for all languages, whereas the MT variant reuses XOR-Retrieve prompt exemplars with translated summaries and queries for 15 languages.\footnote{We were unable to translate our prompt exemplars for 5 languages due to language unavailability in Google Translate: Boro (\texttt{brx}), Garhwali (\texttt{gbm}), Chattisgarhi (\texttt{hne}) and Marwari (\texttt{mwr}). Manipuri (\texttt{mni}) is available in Google Translate in ``Meitei'' script instead of the ``Bengali-Assamese'' script present in the XTREME-UP dataset.}
SWIM-X (120K)$^{MT}$ outperforms the best supervised baseline, mContriever-X$^\heartsuit$ (fine-tuned without MS MARCO) by a huge margin of 12.6 points MRR@10, but performs minimally worse than the MT version by 0.9 points.
Interestingly, none of the evaluated baselines perform well on two extremely low-resource languages, Boro (\texttt{brx}) and Manipuri (\texttt{mni}). 

\subsection{Effectiveness of Summarization in SAP}\label{sec:query-generation-llm} 
\vspace{-1mm}

In our work, we utilize SAP, where we employ extractive summarization as a rationale for PaLM 2 to generate informative multilingual queries. To evaluate the effectiveness of summarization, we assess both models (i.e., contrasting with and without summarization) on cross-lingual retrieval using Recall@5kt on XOR-Retrieve. We additionally evaluate different PaLM 2 model sizes to observe a correlation between retrieval model performance and changes in LLM size, i.e., model parameters. To ensure consistency, we adopt the experimental setup utilized in SWIM-X (500K) for all models. 

Our results are shown in \autoref{fig:cot-vs-standard} (left). we infer two insights: (i) an increase in the LLM size provides diminishing returns in terms of Recall@5kt on XOR-Retrieve. (ii) SAP outperforms standard prompting by at least 0.6 points consistently with all various PaLM-2 generators on XOR-Retrieve, with a maximum improvement of up to 3.2 points Recall@5kt. We observe that PaLM 2 with large sizes (sizes $>$ S) are inherently able to generate coherent queries, leading to diminishing improvements in SAP versus standard prompting.

\subsection{How much Synthetic Data to Generate?}\label{sec:training-data-ablation}
We analyze the optimal amount of synthetic training data required for fine-tuning SWIM-X. \autoref{fig:swim-training-data} depicts the relative improvement in Recall@5kt on XOR-Retrieve. SWIM-X performance (gradually increasing) starts to saturate after 500K synthetic training pairs. The first observation is that with only 2K training pairs, SWIM-X (2K) achieves 49.1 Recall@5kt on XOR-Retrieve, already outperforming the best zero-shot (English-only) baseline. The break-even point occurs at 200K pairs, where SWIM-X (250K) achieves 60.5, outperforming mContriever-X, which achieves a 59.6 Recall@5kt on XOR-Retrieve.

\subsection{Indo-European Language Transferability}
\begin{figure}[t]
    \centering
    \begin{center}
        \includegraphics[trim=0 5 80
        18,clip,width=\textwidth]{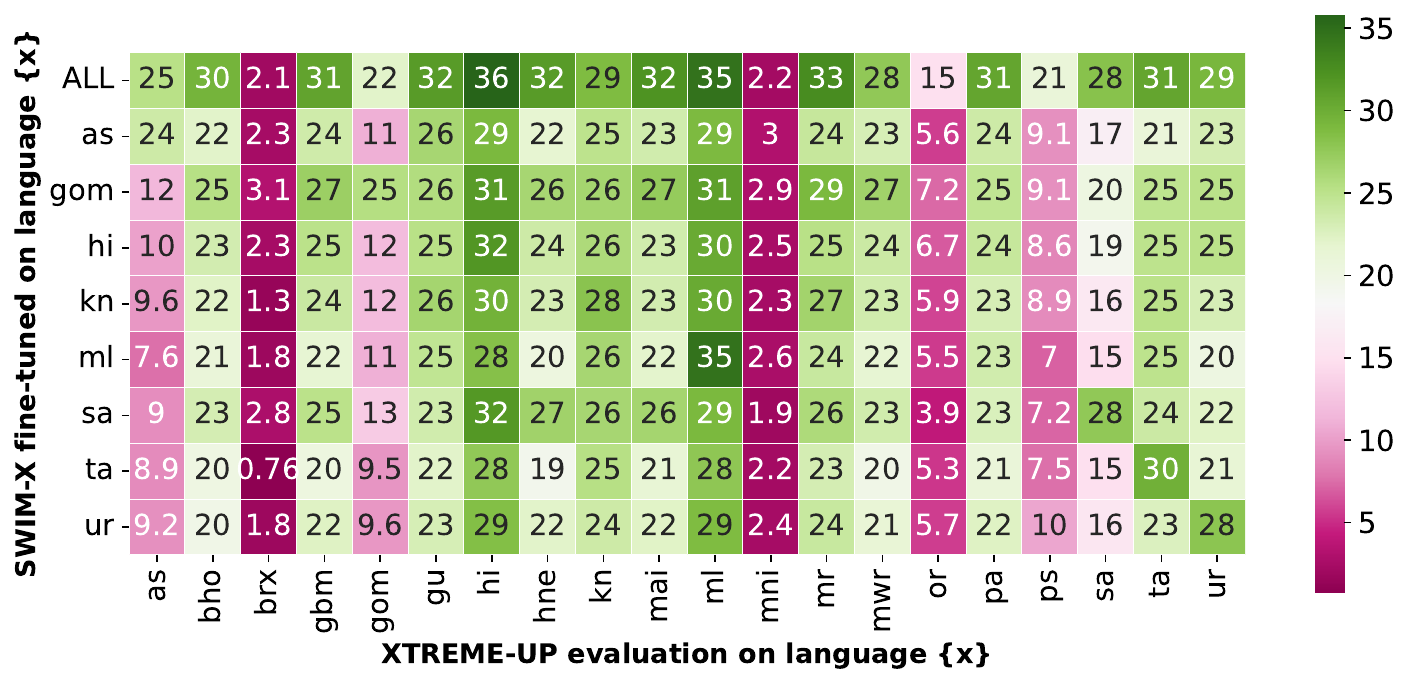}
        \caption{Heatmap showing MRR@10 denoting language-based transfer ability of SWIM-X (120K) across Indo-European languages available in XTREME-UP \cite{ruder2023xtremeup}. (ALL)~denotes SWIM-X fine-tuned on all XTREME-UP languages. \vspace{-1mm}}
        \label{fig:xteme-up-language-transferability}
        \vspace{-1mm}
    \end{center}
\end{figure}

We investigate the language transfer capabilities of the available Indic split (Indo-European language family) in SWIM-IR. We fine-tune individual SWIM-X models for eight selected languages and evaluate them on XTREME-UP. From \autoref{fig:xteme-up-language-transferability}, we observe that SWIM-X models fine-tuned for Konkani (\texttt{gom}) or Hindi (\texttt{hi}) transfers best on all languages in XTREME-UP (rows 3 and 4), whereas fine-tuning for Tamil (\texttt{ta}) transfers worst overall (row 8). Assamese (\texttt{as}), Konkani (\texttt{gom}), Odia (\texttt{or}), Pashto (\texttt{pa}) and Sanskrit (\texttt{sa}) exhibit the lowest zero-shot capabilities with SWIM-X, thereby highlighting the importance of in-language synthetic data. Hindi (\texttt{hi}), Kannada (\texttt{kn}) and Malayalam (\texttt{ml}) demonstrate good zero-shot transfer capabilities with all Indic languages.

\vspace{-1mm}
\section{Ablation Studies}\label{sec:ablation_studies}
\vspace{-1mm}

\begin{figure}[t]
    \centering
    \begin{center}
        \includegraphics[trim=0 0 0 0,clip,width=0.57
        \textwidth]{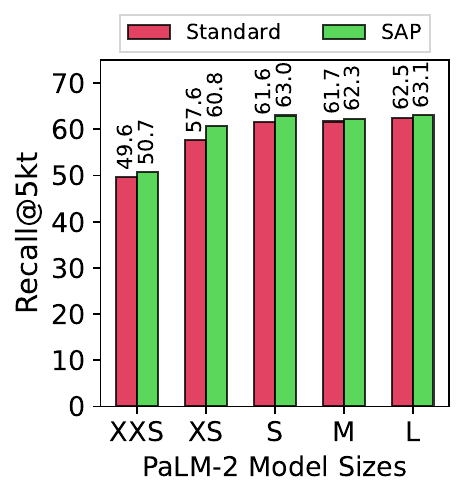}
        \includegraphics[trim=0 0 0 0,clip,width=0.41\textwidth]{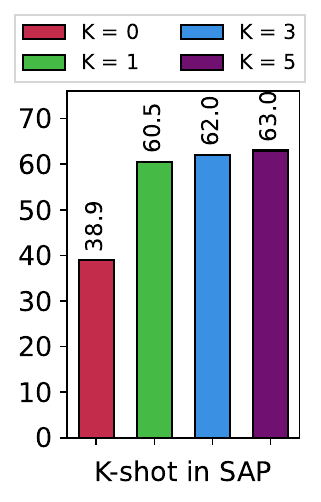}
        \caption{(Left) SAP ({\it Summarize-then-Ask Prompting}) (green) versus standard prompting (red) for various PaLM 2 model sizes. (Right) Varying K-shot prompt exemplars.~SWIM-X is fine-tuned on 500K~SWIM-IR training pairs and evaluated on XOR-Retrieve. \vspace{-2mm}}
        \label{fig:cot-vs-standard}
        \vspace{-2mm}
    \end{center}
\end{figure}

\textbf{Optimal value of k-shot exemplars.} We investigate the optimal value of few-shot exemplars required by PaLM 2 and the variation in the retrieval performance on XOR-Retrieve.\footnote{We limit K = 5 to fit within a context length of 4096 tokens. For additional exemplars, PaLM 2 would need a longer context length increasing the computational cost significantly.} From \autoref{fig:cot-vs-standard} (right), we observe a linear improvement in Recall@5kt with increase in K. Best Recall@5kt is observed with K = 5. SAP is unable to perform well zero-shot (i.e., K = 0) due to the complex nature of the multilingual query generation task which requires few-shot exemplars to understand and generate a summary and a query.

\smallskip
\noindent\textbf{ByT5 tokenizer.}\label{sec:byt5-tokenizer} 
We evaluate whether the poor performance of SWIM-X on low-resource languages in XTREME-UP can be attributed to low-quality language tokenization. 
We replicate SWIM-X using a ByT5-base model as backbone, which contains a language independent tokenizer extension \cite{xue-etal-2022-byt5}. From our results in \autoref{tab:synthetic-query-replacement}, ByT5 models underperform by up to 9.8 points MRR@10 on XTREME-UP, in contrast to mT5-base. Additionally, the performance of SWIM-X on both \texttt{mni} and \texttt{brx} does not improve with ByT5. We leave it as future work to investigate the low performance on \texttt{mni} and \texttt{brx}.

\smallskip
\noindent\textbf{Training split query replacement.} Next, we evaluate the impact of human-generated versus LLM-generated queries on retrieval performance on XTREME-UP. We replace all human-generated queries in the XTREME-UP training split with only synthetic queries generated using PaLM 2 (S). From \autoref{tab:synthetic-query-replacement}, the performance drops by 2.0 points at MRR@10. This confirms that human-generated queries are of better quality, which correlates with an improvement in MRR@10 on XTREME-UP. However, SWIM-X can be fine-tuned efficiently using few synthetic training pairs, by only marginally dropping in retrieval performance.

\vspace{-1mm}
\section{Cost Comparison}\label{sec:cost-comparison}
\begin{figure}[t]
    \centering
    \begin{center}
        \includegraphics[trim=0 0 0 0,clip,width=\textwidth]{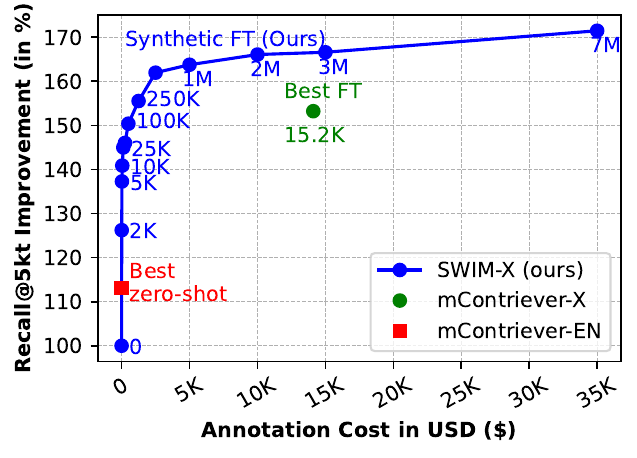}
        \caption{Recall@5kt improvement (in \%) on XOR-Retrieve versus annotation cost in USD (\$) to construct the training dataset. The amount of generated training pairs (human-generated marked in red and green; LLM-generated marked in blue) is provided with each marked data point in the graph.}
        \label{fig:swim-training-data}
    \end{center}
    \vspace*{-\baselineskip}
\end{figure}

\vspace{-1mm}
Generating synthetic training data is relatively inexpensive; however, it is not free. The cost is dependent upon the length of the prompt, input, and output generated from the LLM. The costs also linearly increase with each additional language pair.
At the time of writing, PaLM 2 and similar LLMs cost about 0.0005 USD for 1000 characters in the input and output text.\footnote{PaLM 2 pricing: \href{https://cloud.google.com/vertex-ai/pricing\#generative_ai_models}{cloud.google.com/vertex-ai/pricing}}~Our prompts on average contain about 8--9K characters in the prompt input and generate about 1--2K characters in the output. The relative performance improvement associated with annotation cost in XOR-Retrieve is shown in \autoref{fig:swim-training-data}. Generating 200K synthetic training pairs in SWIM-IR will roughly cost \$1K USD. SWIM-X (200K) performs comparably to the best supervised baseline (mContriever-X), trained on 15.2K human-annotated pairs, requiring roughly 14 times more, i.e., \$14.1K USD to annotate, if we pay an hourly rate of \$18.50 USD per hour for the annotator (local minimum wages is \$11.50 USD/hr) following \cite{miracl}, assuming an estimated annotation cost of 3.0 minutes per example \cite{ruder2023xtremeup}.

\section{Background and Related Work}
The development of pre-trained multilingual LMs has contributed toward recent progress in multilingual retrieval \cite{asai-etal-2021-xor, izacard2022unsupervised, cora, dr-decr, ruder2023xtremeup, miracl, zhang:2022}. Notable baselines in this field include mDPR and mContriever. mDPR \cite{asai-etal-2021-xor, cora, zhang:2022} extends English DPR \cite{karpukhin-etal-2020-dense} to the multilingual setting, while mContriever \cite{izacard2022unsupervised} adopts an unsupervised pre-training objective using the contrastive loss function and data prepared from mC4 \cite{xue2021mt5}, and is fine-tuned on MS MARCO.

\begin{table}[t]
    \centering
    \small    
    \resizebox{\columnwidth}{!}{%
    \setlength\tabcolsep{3pt}
    \begin{tabular}{l c c c c c}
    \toprule
    \textbf{Model} & \textbf{PLM} & \textbf{Query Gen.} & \textbf{\texttt{brx}} & \textbf{\texttt{mni}} & \textbf{MRR@10} \\
    \specialrule{.4pt}{2pt}{0pt}
    \rowcolor{paleaqua} \multicolumn{6}{l}{\textit{1. Models with Byte-level (UTF-8) tokenizer}} \\ 
    mContriever-X$^\heartsuit$ & ByT5 &   Human & 1.8 & 1.0 & 2.1 \\
    SWIM-X (120K)$^{MT}$ & ByT5 & PaLM 2 & 2.1 & 4.9 & 13.3 \\
    SWIM-X (120K) & ByT5 &  PaLM 2 & 5.1 & 5.8 & 15.4 \\
    \rowcolor{paleaqua} \multicolumn{6}{l}{\textit{2. Human-generated query replacement in XTREME-UP}} \\
    mContriever-X$^\heartsuit$ & mT5 &  Human & - & - & 13.5 \\
    SWIM-X ($\approx$10K) & mT5 & PaLM 2 & - & - & 11.5 \\
    \bottomrule
    \end{tabular}}
    \caption{XTREME-UP ablation studies. First, we replace mT5 pre-trained model with ByT5 \cite{xue-etal-2022-byt5}. Next, we replace the human-generated queries in the training dataset with PaLM-2 synthetic queries; MRR@10 scores are macro-averaged for all 20 languages; \texttt{brx} denotes Boro and \texttt{mni} denotes Manipuri. \vspace{-3mm}} 
    \label{tab:synthetic-query-replacement}\vspace{-1mm}
\end{table}

\smallskip
\noindent\textbf{Synthetic data generation.} 
Traditionally, docT5query \cite{nogueira2019doc2query} for query generation has been prominent for generating synthetic training data in English 
\cite{ma-etal-2021-zero,thakur2021beir, wang-etal-2022-gpl, thakur2022domain}. Recently, using LLMs for query generation has gained interest.
\citet{Bonifacio2022InParsUD} proposed InPars, where they few-shot prompt GPT-3 \cite{brown2020language} to generate synthetic queries. Similarly, complementary works \cite{sachan:2022, jeronymo:2023, boystov:2023, falcon:2023, dua:2023} all follow a similar setup as in \citet{Bonifacio2022InParsUD}.
\citet{Dai2022PromptagatorFD} proposed Promptagator, which studied task-dependent few-shot LLM prompting and used the synthetic data for both retrieval and ranking models. Similarly, HyDE \cite{gao2022precise} and GenRead \cite{yu:2023} generate synthetic documents instead of queries. 
However, prior work has focused on English, with the exception of HyDE. In our work, we robustly investigate how LLMs can be used for improving multilingual retrieval systems.

\smallskip
\noindent\textbf{Multilingual datasets.}
Prior work investigates techniques to build multilingual datasets for better fine-tuning or evaluation of dense retrieval models. Datasets such as NeuCLIR \cite{lawrie2023overview}, MKQA \cite{longpre-etal-2021-mkqa} have been constructed using human annotators. Similarly, mMARCO \citep{bonifacio2022mmarco} has been generated using machine translation of MS MARCO~\cite{msmarco}. However, as translated documents are not written by native speakers, mMARCO and similar datasets suffer from artifacts such as ``Translationese''~\cite{clark-etal-2020-tydi}. A concurrent work, JH-POLO \citep{mayfield2023synthetic}, prompts GPT-3 to generate English queries from language specific passages in NeuCLIR.

\section{Discussion and Future Work}
A large-scale construction of SWIM-IR is challenging. Conducting SAP-based LLM generation at a large scale would require an efficient solution. Currently, we support a total of 33 languages. Extending naively to 100 languages would lead to at least 3 times the cost (fixed cost with every language). Hence, naively increasing more languages is not feasible. Instead, in the future, we can focus on generating synthetic data for diverse languages present within groups or clusters, based on linguistic characteristics within a language family or sub-family \cite{rijhwani:2019} and rely on cross-lingual transfer for the remaining languages.

\section{Conclusion}

In this work, we present SWIM-IR, a synthetic multilingual retrieval training dataset with 28 million training pairs across 33 diverse languages. SWIM-IR allows synthetic fine-tuning of multilingual dense retrieval models cheaply without human supervision. SWIM-IR is constructed using SAP, which stands for {\it summarize-then-ask prompting}, assisting the LLM to identify the relevant sections of the input passage, improving the quality of the generated multilingual query.

Our rigorous evaluation across three multilingual retrieval benchmarks assesses our dataset quality. We find that SWIM-X, fine-tuned on SWIM-IR (keeping model training parameters unchanged) outperforms the best supervised cross-lingual baseline by 7.1 points Recall@5kt on XOR-Retrieve and 11.7 points MRR@10 on XTREME-UP, while remaining competitive in monolingual retrieval on MIRACL.

\section{Limitations of SWIM-IR}

SWIM-IR, like any other dataset, is not perfect and has limitations. These limitations do not directly affect the downstream multilingual retrieval task, where dense retrieval models learn how to match relevant passages to queries. The dataset has been created for the ``sole'' purpose of training multilingual retrieval models. We describe below a few noted limitations:

\smallskip
\noindent\textbf{1. Decontextualization.} PaLM 2 captures the salient information from the paragraph, but can generate the query in a reduced context, which cannot be answered without the Wikipedia paragraph. 

\smallskip
\noindent\textbf{2. Code-switching.} PaLM 2 can occasionally generate a code-switched query with words combined from English and the target language. Code-switching is more frequently observed for cross-lingual generation in low-resource languages. 

\smallskip
\noindent\textbf{3. Passage quality and length.} A good quality passage contains relevant information about a topic, which PaLM 2 uses to generate a synthetic query. However, if the passage is really short with little or no information, or contains noisy information, this can likely generate a subpar query. 

\smallskip
\noindent\textbf{4. Factual inconsistencies in LLM generation.} LLMs have been found to generate text lacking sufficient grounding to knowledge sources \cite{dziri-etal-2022-origin, 10.1145/3571730}, thereby posing risks
of misinformation and hallucination in their generated outputs \cite{maynez-etal-2020-faithfulness, raunak-etal-2021-curious,muller2023evaluating}. Queries in SWIM-IR are relevant for the input passage, but are not human-verified, thereby queries may contain factual inconsistencies. We leave it as future work to investigate techniques to improve factual consistency of generated queries \cite{sun-etal-2021-shot, huang-etal-2023-zero}.

\section*{Acknowledgements}
We would like to thank Jinhyuk Lee and other internal reviewers from Google for reviewing our paper and giving feedback on the draft.

\bibliography{naacl_2024}
\clearpage

\appendix
\section{Appendix}
The following supplementary sections in SWIM-IR are arranged as follows:
\begin{compactitem}
    \item \autoref{sec:dataset-release} provides information on the SWIM-IR dataset release.
    \item \autoref{sec:SWIM-IR-extra-material} provides the additional material with SWIM-IR, including the data card, examples, and prompts.
    \item \autoref{sec:content-filtering} provides details on SWIM-IR content filtration.
    \item \autoref{sec:additional-training-details} provides information in detail on hyperparameter tuning and training methodology for baseline models, including multilingual pre-training, synthetic fine-tuning, and passage sampling strategies.
    \item \autoref{sec:SWIM-IR-evaluation} provides statistics for three multilingual retrieval evaluation datasets: XOR-Retrieve, MIRACL, and XTREME-UP. 
    \item \autoref{sec:SWIM-IR-eval-results} contains additional experimental results on XOR-Retrieve and MIRACL. 
\end{compactitem}

\section{Details on SWIM-IR Dataset Release}\label{sec:dataset-release}

\textbf{Dataset release format.} The SWIM-IR dataset will be released and available in multiple formats.
Officially, the dataset is released within the Google Cloud Storage (GCS) cloud storage bucket.\footnote{\href{http://storage.googleapis.com/gresearch/SWIM-IR/swim_ir_v1.tar.gz}{storage.googleapis.com/gresearch/swim\-ir/swim\_ir\_v1.tar.gz}} Later, for longer term preservation, the dataset will be maintained through a TensorFlow Dataset (TFDS). To enable a wider audience within the research community, we plan to release an official copy of SWIM-IR as a Hugging Face dataset \cite{lhoest-etal-2021-datasets}. 

\smallskip
\noindent
\textbf{High quality check.} The SWIM-IR dataset has undergone a high-quality check and a thorough review internally at Google to avoid inaccurate or misleading conclusions drawn from the dataset. High-quality checks are integral to the scientific process to enable researchers to address errors, inconsistencies and identify potential sources of bias within datasets \cite{datacard}. This enables a robust and trustworthy scientific analysis within the community. 

\smallskip
\noindent\textbf{Long term preservation.} SWIM-IR will be available for a long time by continually updating the Tensorflow dataset (TFDS) and Hugging Face dataset. The authors will be responsible for maintaining the dataset and extending the work in the future to support more languages \cite{joshi-etal-2020-state}. Another useful feature is (EN$\rightarrow$L) cross-language retrieval setting, i.e., English query retrieves language-specific passages within a corpus. 

\smallskip
\noindent\textbf{Licensing.} The SWIM-IR corpora is based on multilingual Wikipedia. Therefore for licensing SWIM-IR, we follow the same license as Wikipedia: Creative Commons Attribution-ShareAlike 4.0 Unported License (CC BY-SA 4.0).\footnote{ \href{https://creativecommons.org/licenses/by-sa/4.0/}{https://creativecommons.org/licenses/by-sa/4.0}} The license allows both researchers and industry alike to access the SWIM-IR dataset, copy, and redistribute it for future work.

\section{SWIM-IR Extra Material}\label{sec:SWIM-IR-extra-material} 

\subsection{SWIM-IR Data Card} 
We release the data card associated with the SWIM-IR. The data card was generated using the template provided by the Data Cards Playbook \cite{datacard}. It has been formatted using Markdown.\footnote{The Markdown format and the template are available here: \href{https://github.com/pair-code/datacardsplaybook}{https://github.com/pair-code/datacardsplaybook}} The SWIM-IR data card is provided along with our dataset release on the GitHub repository: \url{https://github.com/google-research-datasets/SWIM-IR}.

\subsection{SWIM-IR Dataset Statistics}\label{sec:SWIM-IR-dataset-examples} 
\begin{figure*}[t]
    \centering
    \begin{center}
        \includegraphics[trim=0 0 0 0,clip,width=\textwidth]{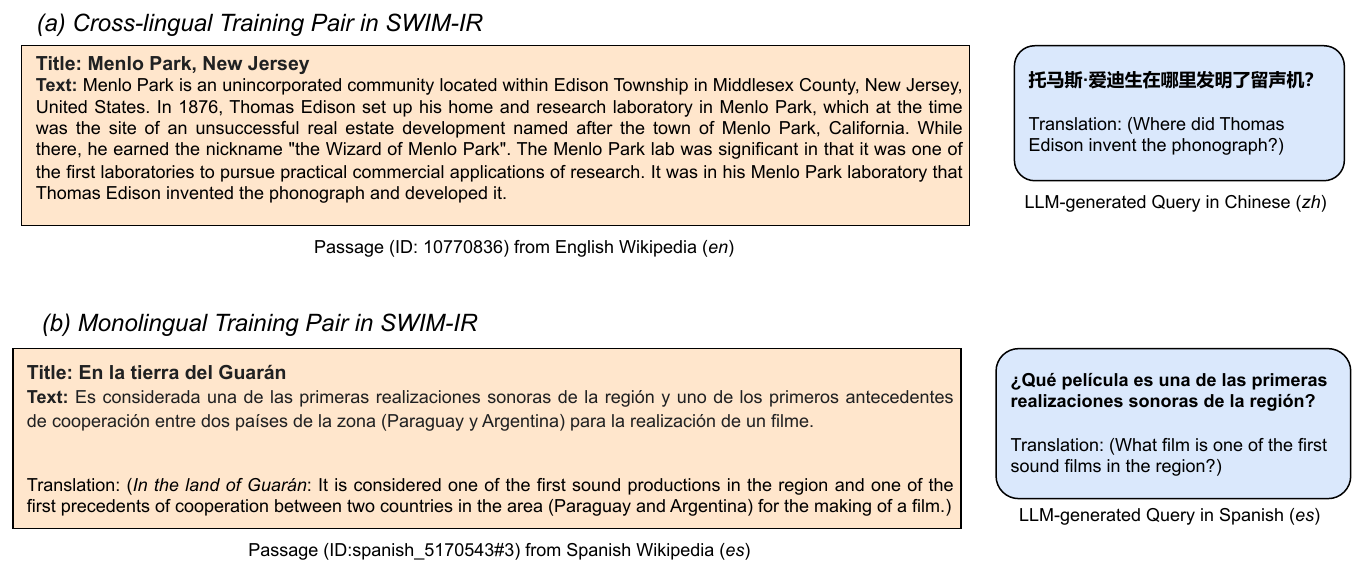}
        \caption{Dataset examples showing both (a) cross-lingual and (b) monolingual training pairs in the SWIM-IR dataset. The passage is selected from English Wikipedia, and PaLM 2 generates the query. A detailed description of all the dataset column headers are provided in Appendix ($\mathsection$\ref{sec:SWIM-IR-dataset-examples}). All translations in the figure above have been provided using Google Translate (\href{https://translate.google.com/}{translate.google.com}) for illustration purposes.}
        \label{tab:swim-ir-examples}
    \end{center}
    \vspace*{-\baselineskip}
\end{figure*}

The languages covered and the amount of training pairs available in SWIM-IR are provided in \autoref{tab:SWIM-IR-dataset-statistics}. The majority of the training pairs (sampled for a maximum of 1M per language pair) are provided for 18 languages in MIRACL, which overlap with the 7 languages in XOR-Retrieve. An additional 100K training pairs come from the rest of the 15 Indo-European languages from XTREME-UP. Two examples from SWIM-IR for each task, cross-lingual and monolingual retrieval, are provided in \autoref{tab:swim-ir-examples}. The cross-lingual example is from Chinese (\texttt{zh}) and the monolingual is from Spanish (\texttt{es}).

Each SWIM-IR training data point has six associated text fields. We describe each field below: 
(i) \texttt{\_id}: denotes the unique identifier of the training pair. (ii) \texttt{title}: denotes the title of the Wikipedia article. (iii) \texttt{text}: denotes the passage extracted from the Wikipedia article. (iv) \texttt{query}: denotes the synthetic multilingual query generated using PaLM 2 \cite{anil2023palm}. (v) \texttt{lang}: denotes the target language in which the query was generated. (vi) \texttt{code}: denotes the ISO code of the generated query language.

\subsection{SWIM-IR Prompts}\label{sec:prompts}
\begin{figure*}[t]
    \centering
    \begin{center}
        \includegraphics[trim=0 0 0 0,clip,width=\linewidth]{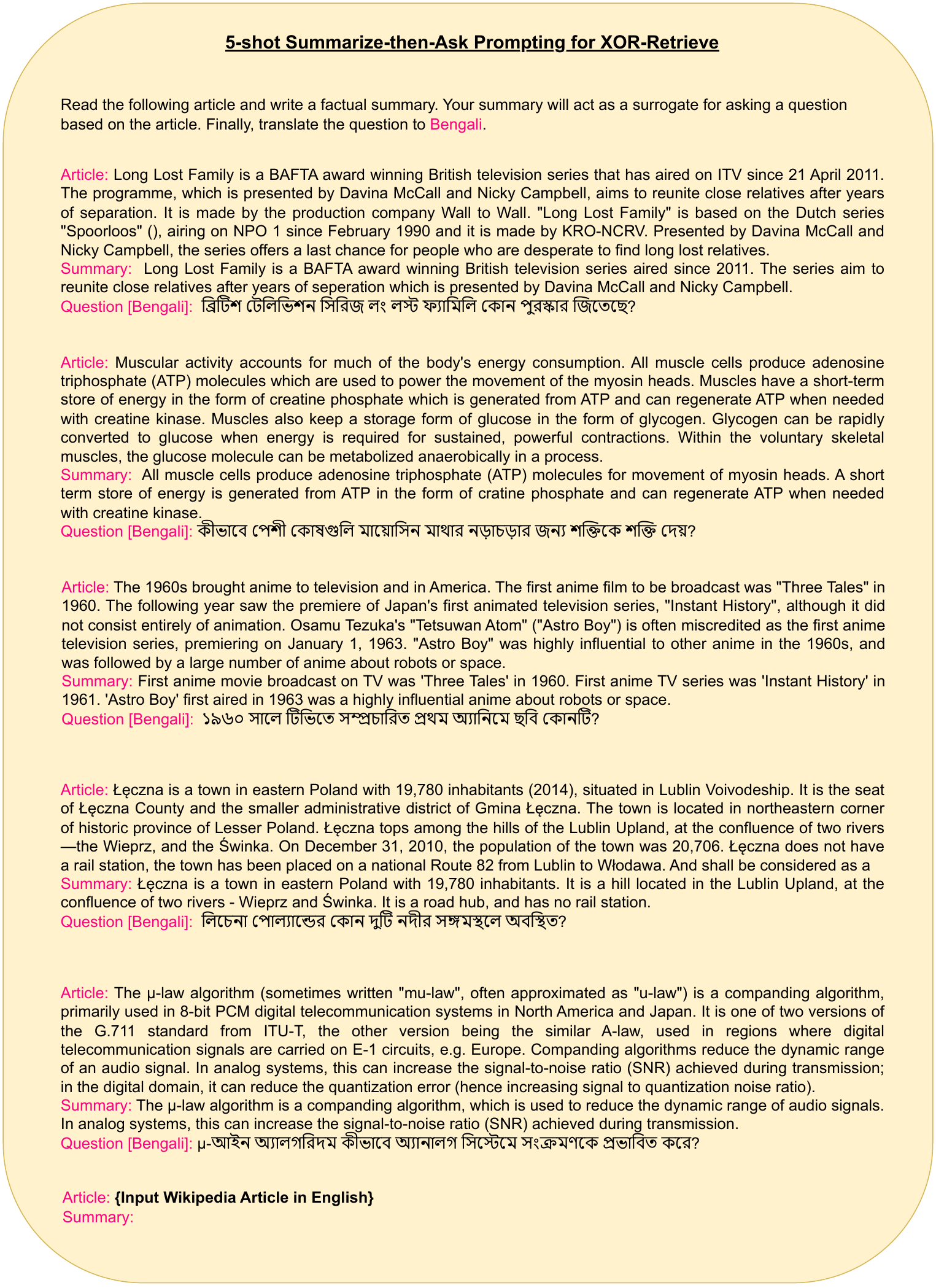}
        \caption{5-shot SAP ({\it Summarize-then-Ask Prompting}) for XOR-Retrieve \cite{asai-etal-2021-xor} is shown for Bengali (\texttt{bn}). There are five exemplars (5-shot) in our cross-lingual query generation task. The passages are randomly selected from the XOR-Retrieve Wikipedia corpus. A summary and a query for all above exemplars is manually written in English by the authors. Finally, the English written query is translated to Bengali (\texttt{bn}) for all above exemplars using Google Translate (\href{https://translate.google.com/}{translate.google.com}).}
        \label{fig:xor-retrieve-prompt}
    \end{center}
    % \vspace*{-\baselineskip}
    \vspace{-1mm}
\end{figure*}

\begin{figure*}[t]
    \centering
    \begin{center}
        \includegraphics[trim=0 300 0 0,clip,width=\linewidth]{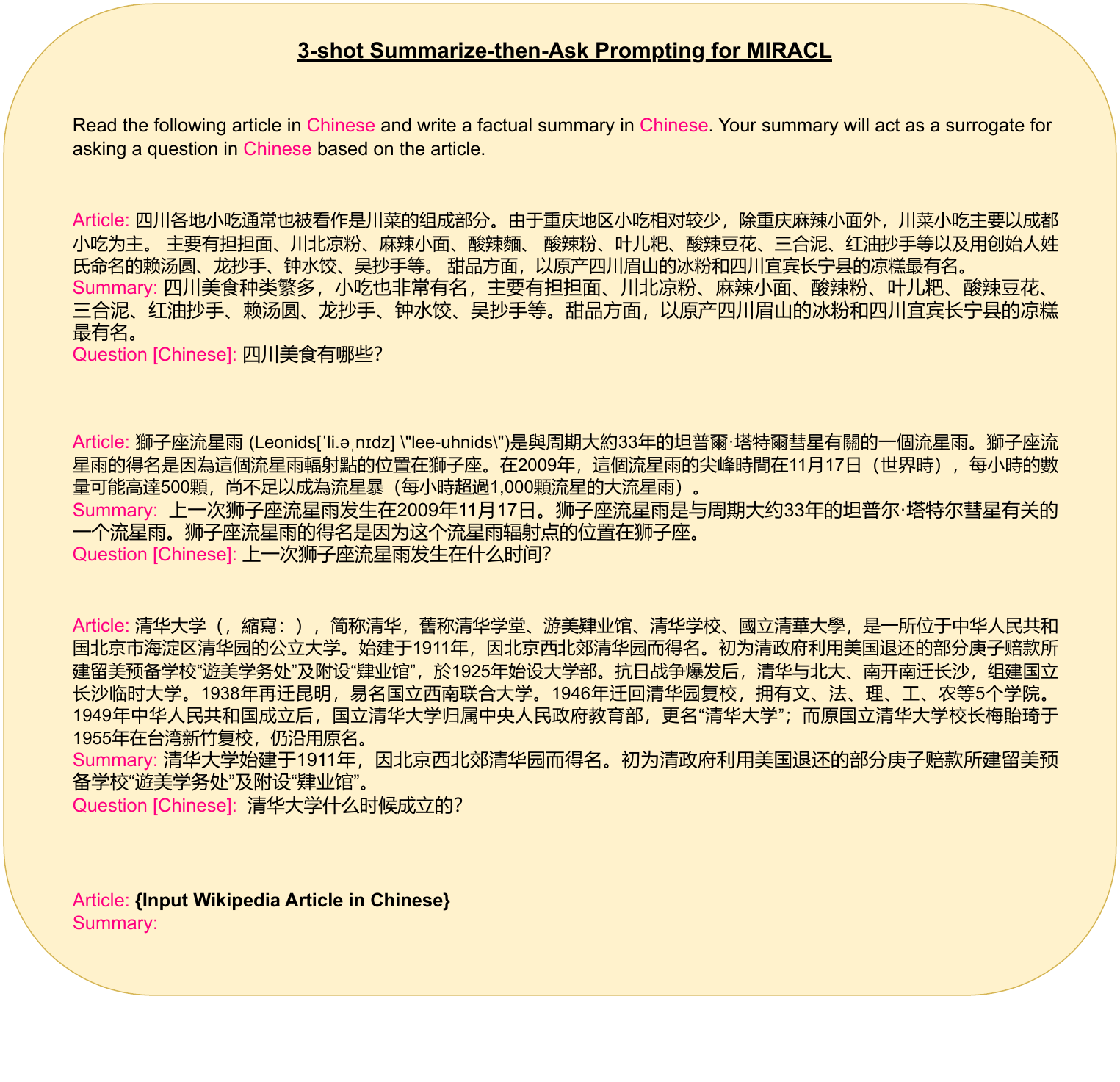}
        \caption{3-shot SAP ({\it Summarize-then-Ask Prompting}) for MIRACL \cite{miracl} is shown for Chinese (\texttt{zh}). There are three exemplars (3-shot) in our monolingual query generation task. The query-passage pairs are randomly selected from MIRACL training set. Finally, the summary for all above exemplars is automatically generated in Chinese (\texttt{zh}) using Google Bard ({\href{https://bard.google.com/}{bard.google.com}}).}
        \label{fig:miracl-prompt}
    \end{center}
    \vspace*{-\baselineskip}
    \vspace{-1mm}
\end{figure*}

\begin{figure*}[t]
    \centering
    \begin{center}
        \includegraphics[trim=0 0 0 0,clip,width=\linewidth]{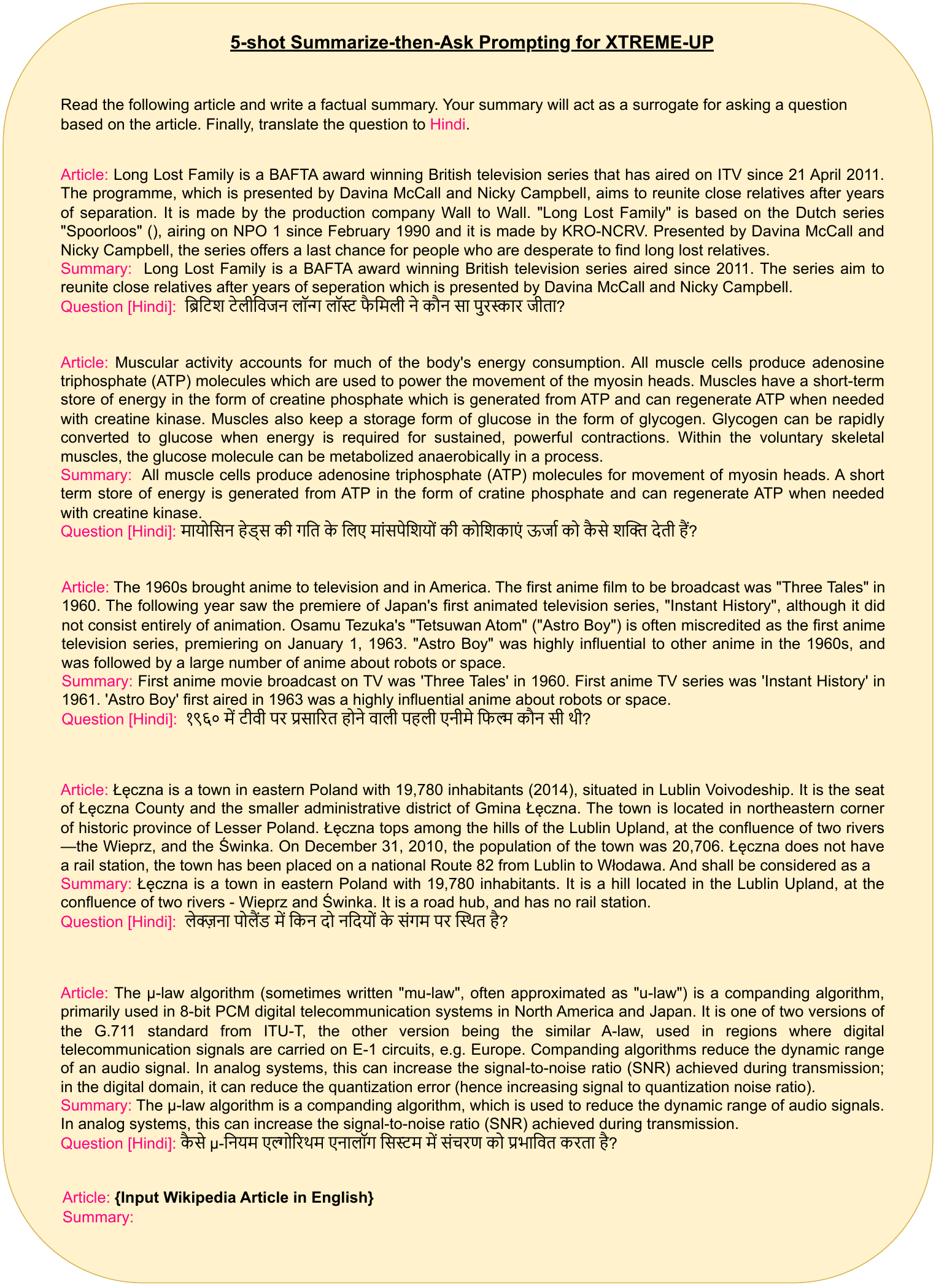}
        \caption{5-shot SAP ({\it Summarize-then-Ask Prompting} with Machine Translation (MT) for XTREME-UP \cite{ruder2023xtremeup} is shown for Hindi (\texttt{hi}). There are five exemplars (5-shot) in our cross-lingual query generation. The exemplars are re-used from XOR-Retrieve. A summary and a query for all above exemplars is manually written in English by the authors. Finally, the English written query is translated to Hindi (\texttt{hi}) for all above exemplars using Google Translate (\href{https://translate.google.com/}{translate.google.com}).}
        \label{fig:xtreme-up-prompt}
    \end{center}
    % \vspace*{-\baselineskip}
    \vspace{-1mm}
\end{figure*}

All prompts and their templates (across all 33 languages) used to develop SWIM-IR are available in the GitHub repository.\footnote{\href{https://github.com/google-research-datasets/SWIM-IR}{https://github.com/google-research-datasets/SWIM-IR}}
We provide a few individual prompt examples for all three datasets in the Appendix: (1) XOR-Retrieve (English passage; synthetic Bengali query) in \autoref{fig:xor-retrieve-prompt}, (2) MIRACL (Chinese passage; synthetic Chinese query) in \autoref{fig:miracl-prompt}, and (3) XTREME-UP (English passage; synthetic Hindi query) in \autoref{fig:xtreme-up-prompt}. 

\vspace{-1mm}
\section{Content Filtration}\label{sec:content-filtering}
\vspace{-1mm}

LLMs have been shown to generate undesirable content, particularly when primed with material aimed at eliciting negative patterns or associations from the model's training data \cite{gehman:2020, bender:2021}. Initially, we expected that the sampled Wikipedia passages would predominantly contain safe material suitable for prompting LLMs. However, after examination, we discovered that between 6--10\% of the pairs contained sensitive subjects and adult content (i.e., weapons; violence and abuse; accidents and disasters; death and tragedy; war and conflict). To address this issue, we used the Google Cloud Natural Language content classification categories\footnote{\href{https://cloud.google.com/natural-language/docs/categories}{cloud.google.com/natural-language/docs/categories}} to identify and remove pairs where either the original sampled passage or the resulting LLM generated query has a content classification of either \texttt{/Adult} or any of the \texttt{/Sensitive Subjects} labels.

\section{Additional Technical Details}\label{sec:additional-training-details}

\subsection{mContriever Pre-training}\label{sec:mc4-pretraining}
In the original implementation of mContriever \cite{izacard2022unsupervised}, the authors initialized the model using the mBERT \cite{devlin-etal-2019-bert} pre-trained language model (PLM). Subsequently, the model was jointly pre-trained on 29 languages covering the CCNet dataset \cite{wenzek2020ccnet} with a contrastive pre-training objective. 

In our adaptation of mContriever, we initialize using the mT5-base model checkpoint \cite{xue2021mt5}. Next, we jointly pre-train the model on 101 languages\footnote{The list of all 101 languages in mC4 can be found at: \href{https://www.tensorflow.org/datasets/catalog/c4\#c4multilingual}{www.tensorflow.org/datasets/catalog/c4}} available in mC4 dataset \cite{xue2021mt5}. For each mC4 document, we sample two random non-overlapping texts with a maximum text span size of 256 tokens. Similar to the mT5 pre-training objective \cite{xue2021mt5}, examples were not uniformly sampled over languages; instead, the probability of selecting a training sample from a specific language is directly proportional to the amount of training data available in the mC4 dataset. We randomly sample a maximum of 20K samples per language and use them as a validation subset.

We optimize our mContriever model with the AdamW optimizer \cite{loshchilov2018decoupled} with a learning rate of $1e^{-3}$, batch size of 8192, and for 600K pre-training steps. During the first 500K pre-training steps, we use a language-mixed training objective, where a single training batch can contain examples across multiple languages. For the subsequent 100K training steps, we use a language-unmixed training objective, where a single training batch contains all examples from only a single language, i.e., no mixing of different language pairs within a training batch. 
We internally conducted a brief evaluation of the mContriever pre-trainining strategies using language-mixing (500K) and with both language-mixing and unmixing (600K) checkpoints. Notably on XOR-Retrieve, we observed a significant performance improvement with the additional language-unmixed pre-training, resulting in an improvement of 7.3 points Recall@5kt.

\subsection{Supervised Baselines}\label{sec:baseline-models}
\textbf{XOR-Retrieve.}~For the zero-shot baseline model, we fine-tune on the English-only MS MARCO \cite{msmarco} dataset using our base initialization model, mT5 \cite{xue2021mt5}. We use in-batch negatives, AdamW optimizer \cite{loshchilov2018decoupled} and with a learning rate of $1e^{-3}$. The query sequence length is set to a maximum sequence length of 64 tokens, whereas the document is limited to a maximum sequence length of 256 tokens. 
On MS MARCO, models are fine-tuned with a batch size of 4096 and for an additional 50K training steps.

For our supervised baselines, we fine-tune on the XOR-Retrieve training dataset containing 15,250 training pairs. Each training pair in XOR-Retrieve is accompanied by one hard negative \cite{asai-etal-2021-xor}. We fine-tune our baseline models on XOR-Retrieve using triplets containing the query, relevant passage and a single hard negative. We use the AdamW optimizer \cite{loshchilov2018decoupled}, a learning rate of $1e^{-3}$, a batch size of 4096 and fine-tune the model for 15K training steps.

\smallskip
\noindent\textbf{MIRACL.}~For the zero-shot baseline model, we first fine-tune on the MS MARCO \cite{msmarco} dataset. We use the same fine-tuning setup as described for XOR-Retrieve. For monolingual supervised models, we use the MIRACL training data. MIRACL authors provides between one to nine hard negatives for each training query. We randomly sample up to a maximum of four hard negatives for each query and use the AdamW optimizer \cite{loshchilov2018decoupled}, learning rate of $1e^{-3}$, a batch size of 4096 and fine-tune the model for 15K training steps.

\smallskip
\noindent\textbf{XTREME-UP.}~For the zero-shot baseline model, we fine-tune on the MS MARCO \cite{msmarco} dataset. For the supervised baselines, we use the XTREME-UP training data containing 13,270 training pairs and fine-tune with only in-batch negatives (i.e., no hard negatives). We use the AdamW optimizer \cite{loshchilov2018decoupled}, a learning rate of $1e^{-3}$, a batch size of 1024, and fine-tune the model for 5K training steps.

\begin{figure}[t]
    \centering
    \begin{center}
        \includegraphics[trim=0 0 0 0,clip,width=\textwidth]{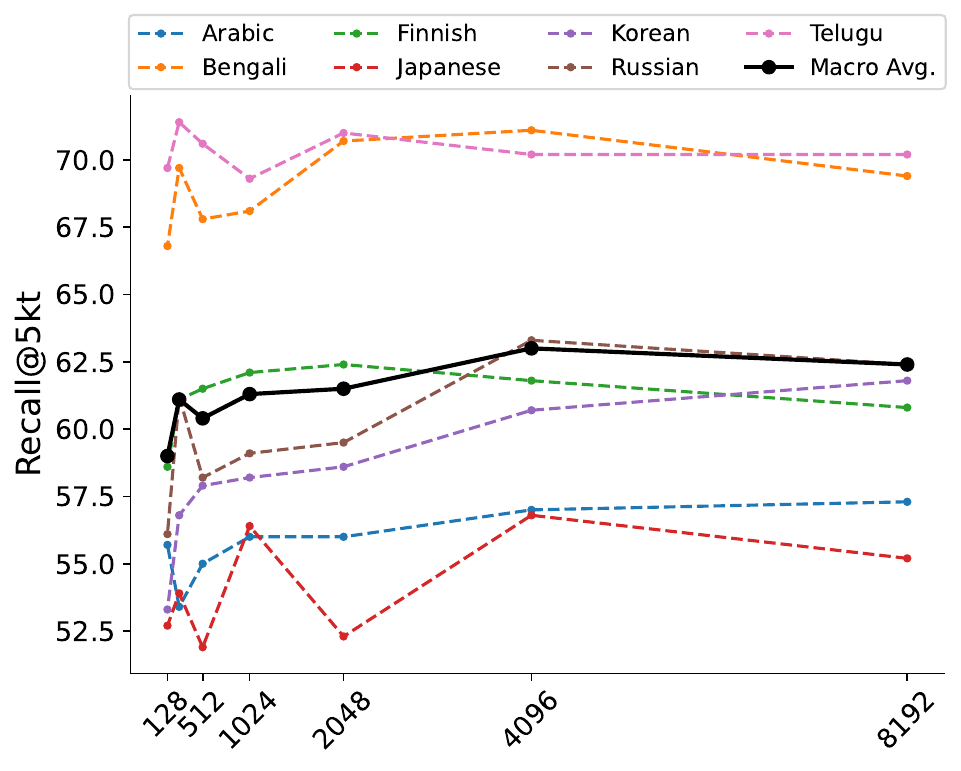}
        \caption{Training batch size ablation of SWIM-X (500K) on XOR-Retrieve \cite{asai-etal-2021-xor}. The best Recall@5kt is achieved with 4096 training batch size. To avoid overfitting, we fine-tune all SWIM-X variants on 500K SWIM-IR training pairs  with decreasing training steps of \{40K, 40K, 30K, 30K, 20K, 20K, 15K\} for increasing batch sizes of \{128, 256, 512, 1024, 2048, 4096, 8192\} respectively.}
        \label{fig:batch-size-ablation}
    \end{center}
    \vspace*{-\baselineskip}
\end{figure}

\subsection{Synthetic Baselines}\label{sec:synthetic-baselines} 
We fine-tune all SWIM-X models using in-batch negatives (no hard negatives), AdamW optimizer \cite{loshchilov2018decoupled} and with a learning rate of $1e^{-3}$. The pre-trained language model for SWIM-X is the mT5-base model with 580M parameters \cite{xue2021mt5}. The batch size and the training steps varies for each dataset. An ablation for batch size is provided in \autoref{fig:batch-size-ablation}. Training data is evenly distributed across all languages present. For example, if there are 100K pairs with 5 different languages, each language contains around 20K training pairs. 

\smallskip
\noindent\textbf{XOR-Retrieve.} SWIM-X is fine-tuned with a batch size of 4096 and with a maximum of 50K training steps on SWIM-IR~cross-lingual pairs. For the 500K training pairs, we fine-tune for 20K steps, and for the maximum of 7M pairs, we fine-tune for 50K training steps. The training pairs within a single batch include language-mixing, i.e., one or more language-specific training pairs are sampled within a single training batch.

\smallskip
\noindent\textbf{MIRACL.} SWIM-X is fine-tuned for a batch-size of 4096 and for a maximum of 15K training steps on SWIM-IR monolingual pairs. As shown in \cite{roy:2020, zhang:2022}, language-unmixed training setup is shown to work well for monolingual retrieval. Following prior work, SWIM-X training pairs include language unmixing, i.e., all pairs are from a single language. The examples are uniformly sampled across all languages, i.e., probability that a training sample comes from a specific language is equal for all languages, unlike during mC4 pre-training. 

\smallskip
\noindent\textbf{XTREME-UP.} SWIM-X is fine-tuned for a batch size of 1024 and for a maximum of 15K training steps on SWIM-IR~cross-lingual (Indic) pairs. Similar to XOR-Retrieve, the training pairs include language-mixing within a single batch during SWIM-X fine-tuning.

\subsection{Stratified Sampling in SWIM-IR}\label{sec:sampling-strategy}
In our work, we use a stratified sampling technique to select a subset of passages from the Wikipedia corpus,  aiming for a relatively uniform distribution of training samples across all languages. Our Wikipedia corpus contains entities which are sorted alphabetically (A-Z). We then compute inclusion threshold $I_{th}$, which is defined as $I_{th} = D_{sample} / D_{total}$, where $(D_{sample})$ is number of passages required to sample and $(D_{total})$ is the total numbers of passages in corpus.
Next, for each passage ($p_i$) in the corpus, we randomly generate an inclusion probability $\hat{p_i} \in [0,1]$. We select the passage ($p_i$) if $p_i \leq I_{th}$. This approach ensures a uniform sampling of passages with Wikipedia entities between all letters (A-Z).\footnote{All Wikipedia entities starting with a non-alphabet are included in the beginning of the Wikipedia corpus.}

\section{Evaluation Dataset Information}\label{sec:SWIM-IR-evaluation} 

We evaluate on three multilingual retrieval benchmarks: (i) \textbf{XOR-Retrieve} \cite{asai-etal-2021-xor}, (ii) \textbf{MIRACL} \cite{miracl} and (iii) \textbf{XTREME-UP} \cite{ruder2023xtremeup}. We excluded NeuCLIR \cite{lawrie2023overview} from our evaluation as it contained a fewer subset of languages namely, Chinese (\texttt{zh}), Farsi (\texttt{fa}) and Russian (\texttt{ru}). Although MKQA \cite{longpre-etal-2021-mkqa} contained a wider variety of languages, it is primarily used for question-answering (QA) rather than multilingual retrieval. All three selected evaluation datasets contain a training split. Only XTREME-UP has released their test split publicly, which we use for evaluation in the paper. However, for both XOR-Retrieve and MIRACL, we evaluate on the development split.

\textbf{XOR-Retrieve} \cite{asai-etal-2021-xor} is a cross-lingual open retrieval training and evaluation task within \textsc{TyDi}-QA \cite{clark-etal-2020-tydi}. 
XOR-Retrieve contains 15K human annotated relevant passage-query pairs in the training set with one hard negative and 2K passage-answer pairs in the {\it dev} set. 
The corpus $C$ contains 18.2M passages with a maximum of 100 word tokens from the English Wikipedia. The queries are multilingual and cover seven languages.
We evaluate our models using recall at m kilo-tokens, i.e., Recall@{m}kt, which computes the fraction of queries for which the minimal answer is contained within the top $m$ thousand tokens of the retrieved passages.
Following prior work in \citet{asai-etal-2021-xor}, we evaluate our models at Recall@5kt and Recall@2kt.

\smallskip
\noindent\textbf{MIRACL} \cite{miracl} is a monolingual open retrieval evaluation task containing 18 languages. 
MIRACL was developed on top of Mr.~\textsc{TyDi} \cite{zhang-etal-2021-mr}, and covers more languages and provides denser judgments by human annotators. 
The test set is not publicly released, hence in this paper we evaluate using the dev set. 
The training set contains 88,288 pairs, with the exception of Yoruba (\texttt{yo}) and German (\texttt{de}) which do not have any training data available. The authors also provide labeled hard negatives for the training query-passage pairs. 
The dev set contains around 13,495 query-passage pairs. 
The corpus $C$ in MIRACL are language-specific Wikipedia articles with various sizes starting from smallest, Yoruba (\texttt{yo}) with 49K passages, till the largest, English (\texttt{en}) with 39.2M passages. Following prior work in \citet{miracl} and \citet{kamalloo2023evaluating}, we evaluate our models at nDCG@10 and Recall@100.

\smallskip
\noindent\textbf{XTREME-UP} \citet{ruder2023xtremeup} contains diverse information-access and user-centric tasks focused on under-represented languages. In our work, we evaluate a cross-lingual retrieval task containing 5,280 query-passage pairs in the training set. The corpus $C$ contains 112,426 passages sampled from \textsc{TyDi}-QA \cite{clark-etal-2020-tydi}. The test set contains 10,705 query-passage pairs for evaluation. The cross-language retrieval for the question-answering (QA) task contains 20 under-represented Indic languages. Following prior work in \citet{ruder2023xtremeup}, we evaluate our models at MRR@10.

\section{Additional Results}\label{sec:SWIM-IR-eval-results}

\smallskip
\noindent\textbf{XOR-Retrieve.}~In \autoref{tab:xor-retrieve-results-supp}, we report the Recall@2kt scores across all multilingual retrievers on XOR-Retrieve. We find similar trends for improvement, SWIM-X (7M) outperforms the best supervised model, mContriever-X, by 3.9 points at Recall@2kt. The SWIM-X (7M) without mC4 pre-training is a strong baseline outperforming SWIM-X (7M) with mC4 pre-training on 4 out of the 7 languages evaluated in XOR-Retrieve. 

\smallskip
\noindent\textbf{MIRACL.}~In \autoref{tab:miracl-results-supp}, we report the Recall@100 scores across all multilingual retrievers on MIRACL. mContriever-X achieves the highest Recall@100 score of 86.5, SWIM-X on the other hand achieves 78.9 at Recall@100, which is competitive and outperforms both the zero-shot baselines, i.e., mDPR-EN and mContriever-EN. For Yoruba, Our SWIM-X outperforms the supervised mContriever-X which shows the importance of synthetic training data for low-resource languages, as the MIRACL supervised training dataset does not contain training pairs in Yoruba (i.e., no human-labeled training pairs). 

\begin{table*}[t!]
\centering
\small
\resizebox{\textwidth}{!}{
\begin{tabular}{c c@{\hskip 0.3in} c c@{\hskip 0.3in} c c}
\toprule
\multicolumn{2}{c}{\textbf{Cross-Lingual (18)}} & \multicolumn{2}{c}{\textbf{Monolingual (18)}} & \multicolumn{2}{c}{\textbf{Cross-Lingual (15)}} \\
Q-P Lang. & \#~Train Pairs & Q-P Lang. & \#~Train Pairs & Q-P Lang. & \#~Train Pairs \\ \specialrule{.4pt}{2pt}{0pt}
\rowcolor{paleaqua} \multicolumn{4}{c}{MIRACL \cite{miracl}} & \multicolumn{2}{c}{XTREME-UP \cite{ruder2023xtremeup}}\\
\texttt{ar}-\texttt{en} & 901,363 & \texttt{ar}-\texttt{ar} &  890,389 & \texttt{as}-\texttt{en} & 5,899 \\ 
\texttt{bn}-\texttt{en} & 909,748 & \texttt{bn}-\texttt{bn} & 257,327 & \texttt{bho}-\texttt{en} & 5,763 \\ 
\texttt{de}-\texttt{en} & 909,145 & \texttt{de}-\texttt{de} &  943,546 & \texttt{gom}-\texttt{en} & 5,755 \\ 
\texttt{en}-\texttt{en} &    -    & \texttt{en}-\texttt{en} &  936,481 & \texttt{gu}-\texttt{en} & 5,870 \\ 
\texttt{es}-\texttt{en} & 905,771 & \texttt{es}-\texttt{es} &  947,340 & \texttt{kn}-\texttt{en} & 5,763 \\ 
\texttt{fa}-\texttt{en} & 910,295 & \texttt{fa}-\texttt{fa} &  973,409 & \texttt{mai}-\texttt{en} & 5,768 \\ 
\texttt{fi}-\texttt{en} & 906,429 & \texttt{fi}-\texttt{fi} &  967,139 & \texttt{ml}-\texttt{en} & 5,907 \\ 
\texttt{fr}-\texttt{en} & 911,694 & \texttt{fr}-\texttt{fr} &  977,900 & \texttt{mni}-\texttt{en} & 5,604 \\ 
\texttt{hi}-\texttt{en} & 919,729 & \texttt{hi}-\texttt{hi} & 466,272 & \texttt{mr}-\texttt{en} & 5,977 \\ 
\texttt{id}-\texttt{en} & 907,826 & \texttt{id}-\texttt{id} &  837,459 & \texttt{or}-\texttt{en} & 5,837 \\ 
\texttt{ja}-\texttt{en} & 906,862 & \texttt{ja}-\texttt{ja} &  893,520 & \texttt{pa}-\texttt{en} & 5,840 \\ 
\texttt{ko}-\texttt{en} & 905,669 & \texttt{ko}-\texttt{ko} &  941,459 & \texttt{ps}-\texttt{en} & 5,694 \\ 
\texttt{ru}-\texttt{en} & 904,933 & \texttt{ru}-\texttt{ru} &  915,693 & \texttt{sa}-\texttt{en} & 5,779 \\ 
\texttt{sw}-\texttt{en} & 905,242 & \texttt{sw}-\texttt{sw} & 123,099 & \texttt{ta}-\texttt{en} & 5,930 \\ 
\texttt{te}-\texttt{en} & 902,190 & \texttt{te}-\texttt{te} & 220,431 & \texttt{ur}-\texttt{en} & 5,816 \\ 
\texttt{th}-\texttt{en} & 914,610 & \texttt{th}-\texttt{th} & 451,540 \\ 
\texttt{yo}-\texttt{en} & 902,467 & \texttt{yo}-\texttt{yo} & 43,211 \\
\texttt{zh}-\texttt{en} & 921,701 & \texttt{zh}-\texttt{zh} & 946,757 \\ \midrule
% Total & 15,532,876 & Total & 12,732,972 \\ \midrule
\multicolumn{6}{c}{Overall Training Pairs = 28,265,848} \\
        \bottomrule
    \end{tabular}
    \caption{Dataset Statistics of SWIM-IR across both cross-lingual and monolingual settings; (Q-P Lang.) denotes the language code of the query-passage training pair in SWIM-IR; (\#~Train Pairs) denotes the count of the relevant training pairs containing the synthetic query and original passage pair.}
    \label{tab:SWIM-IR-dataset-statistics}
    }
\end{table*}

\clearpage

\begin{table*}[t]
    \centering
    \small    
    \resizebox{\columnwidth}{!}{%
    \setlength\tabcolsep{3.5pt}
    \begin{tabular}{l|l|c|c|c|ccccccc}
       \toprule
       Model & PLM & PT & Finetune & \multicolumn{8}{c}{\textbf{Recall@2kt}} \\
       & & & (Datasets) & Avg. & Ar & Bn & Fi & Ja & Ko & Ru & Te \\ \specialrule{.4pt}{2pt}{0pt}
       \rowcolor{paleaqua} \multicolumn{12}{l}{\textit{Existing Supervised Baselines (Prior work)}} \\
       Dr. DECR \cite{dr-decr} & XLM-R & WikiM & \scriptsize{NQ + XOR$^*$} & 66.0 & -- & -- & -- & -- & -- & -- & -- \\ 
       mDPR \cite{asai-etal-2021-xor} & mBERT & --- & \scriptsize{XOR} & 40.5 & 38.8 & 48.4 & 52.5 & 26.6 & 44.2 & 33.3 & 39.9 \\ 
       mBERT + xQG \cite{Zhuang2023AugmentingPR} & mBERT & --- & \scriptsize{XOR} & 46.2 & 42.4 & 54.9 & 54.1 & 33.6 & 52.3 & 33.8 & 52.5 \\ \midrule
       Google MT + DPR \cite{asai-etal-2021-xor} & BERT & --- & \scriptsize{NQ} & 62.2 & 62.5 & 74.7 & 57.3 & 55.6 & 60.0 & 52.7 & 72.3 \\
       OPUS MT + DPR \cite{asai-etal-2021-xor} & BERT & --- & \scriptsize{NQ} & 42.7 & 43.4 & 53.9 & 55.1 & 40.2 & 50.5 & 30.8 & 20.2 \\ 
       \rowcolor{paleaqua} \multicolumn{12}{l}{\textit{Zero-shot baselines (English-only supervision)}} \\
       mContriever & mT5 & mC4 & --- &  29.9 & 27.2 & 23.0 & 35.0 & 27.0 & 27.7 & 35.0 & 34.0 \\
       mDPR-EN  & mT5 & --- & \scriptsize{MS MARCO} & 30.6 & 26.2 & 26.0 & 37.9 & 32.8 & 24.6 & 34.6 & 32.4 \\ 
       mContriever-EN & mT5 & mC4 & \scriptsize{MS MARCO} & 33.8 & 27.8 & 24.3 & 42.4 & 29.9 & 31.2 & 40.5 & 40.3 \\ 
       \rowcolor{paleaqua} \multicolumn{12}{l}{\textit{Supervised Baselines (Cross-lingual supervision)}} \\
       mDPR-X  & mT5 & --- & \scriptsize{XOR} & 43.6 & 43.7 & 50.0 & 44.6 & 36.1 & 41.1 & 35.9 & 54.2 \\
       mContriever-X & mT5 & mC4 & \scriptsize{XOR} & 46.6 & 40.1 & 62.5 & 47.1 & 38.2 & 44.2 & 38.4 & 55.5 \\
       \specialrule{.4pt}{0pt}{2pt}
       mDPR-X & mT5 & --- & \scriptsize{MS MARCO + XOR} & 49.5 & 46.0 & 63.8 & 49.0 & 39.0 & 48.4 & 43.9 & 56.3 \\ 
       mContriever-X & mT5 & mC4 & \scriptsize{MS MARCO + XOR} & 53.0 & 47.6 & 65.1 & 51.6 & 47.3 & 50.2 & 44.3 & 65.1 \\ 
       \rowcolor{paleaqua} \multicolumn{12}{l}{\textit{Synthetic Baselines (Our work)}} \\
       SWIM-X (500K) & mT5 & --- & \scriptsize{SWIM-IR} & 49.2 & 46.3 & 57.2 & 49.0 & 42.7 & 45.6 & 44.7 & 58.8  \\
        SWIM-X (500K) & mT5 & mC4 & \scriptsize{SWIM-IR} & 53.3 & 46.6 & 61.8 & 51.9 & 46.5 & 49.1 & 55.3 & 61.8 \\
       \specialrule{.4pt}{0pt}{2pt}
SWIM-X (7M) & mT5 & --- & \scriptsize{SWIM-IR} & 56.6 & 50.8 & 65.1 & 56.1 & 48.1 & 54.0 & 55.7 & 66.4 \\ 
SWIM-X (7M) & mT5 & mC4 & \scriptsize{SWIM-IR} & 56.9 & 53.4 & 67.8 & 55.1 & 49.4 & 52.6 & 55.3 & 64.7 \\
       \bottomrule
    \end{tabular}}
    \caption{Experimental results showing Recall@2kt for cross-lingual retrieval on XOR-Retrieve dev \cite{asai-etal-2021-xor}; (PLM) denotes the pre-trained language model; (PT) denotes the pre-training dataset; ($^*$) Dr.DECR is fine-tuned in a complex training setup across more datasets (\S\ref{sec:implementation-details}); WikiM denotes WikiMatrix \cite{schwenk-etal-2021-wikimatrix}; XOR denotes XOR-Retrieve; SWIM-X (ours) is fine-tuned on 500K and 7M synthetic data.}
    \label{tab:xor-retrieve-results-supp}
    % \vspace{-0.5cm}
\end{table*}
\begin{table*}[]
\centering
\resizebox{\textwidth}{!}{
\begin{tabular}{l|c|cccccccccccccccccc}
\toprule
 Model & Avg. & \texttt{ar} & \texttt{bn} & \texttt{en} & \texttt{es} & \texttt{fa} & \texttt{fi} & \texttt{fr} & \texttt{hi} & \texttt{id} & \texttt{ja} & \texttt{ko} & \texttt{ru} & \texttt{sw} & \texttt{te} & \texttt{th} & \texttt{zh} & \texttt{de} & \texttt{yo} \\
\specialrule{.4pt}{2pt}{0pt}
\rowcolor{paleaqua} \multicolumn{20}{l}{\textit{Existing Supervised Baselines (Prior work)}} \\ 
BM25 & 77.2 & 88.9 & 90.9 & 81.9 & 70.2 & 73.1 & 89.1 & 65.3 & 86.8 & 90.4 & 80.5 & 78.3 & 66.1 & 70.1 & 83.1 & 88.7 & 56.0 & 57.2 & 73.3 \\
mDPR & 79.0 & 84.1 & 81.9 & 76.8 & 86.4 & 89.8 & 78.8 & 91.5 & 77.6 & 57.3 & 82.5 & 73.7 & 79.7 & 61.6 & 76.2 & 67.8 & 94.4 & 89.8 & 79.5 \\
Hybrid & 88.0 & 94.1  &93.2  & 88.2  &94.8  & 93.7 & 89.5 & 96.5 & 91.2 & 76.8 & 90.4 & 90.0 &87.4  & 72.5 & 85.7 & 82.3 & 95.9 &88.9 & 80.7 \\
Cohere-API & 76.9 & 85.4 & 85.6 & 74.6 & 71.7 &  77.1&  80.9&  81.6&  72.4&  68.3&  81.6&  77.1& 76.7 &  66.6&  89.8&  86.9&  76.9& 72.5 & 57.6 \\
% \specialrule{.4pt}{2pt}{0pt}
\rowcolor{paleaqua} \multicolumn{20}{l}{\textit{Zero-shot baselines (English-only supervision)}} \\
% \specialrule{.4pt}{0pt}{2pt}
mDPR-EN & 76.9 & 85.5 & 85.9 & 72.4 & 66.8  & 79.7 & 86.0 & 71.4 & 74.2 & 67.0 & 80.1 & 77.1& 77.4  & 80.2 & 91.9 & 84.8 & 68.5& 70.9 & 58.6 \\
mContriever-EN & 76.6 & 73.5 & 80.8 & 52.1 & 49.5  & 61.7 & 66.0 & 51.8 & 50.3 & 63.5 & 65.6 & 56.3& 58.9  & 73.5 & 85.9 & 76.6 & 58.2& 36.3 & 30.2 \\
% \specialrule{.4pt}{2pt}{0pt}
\rowcolor{paleaqua} \multicolumn{20}{l}{\textit{Supervised Baselines (Monolingual supervision)}} \\
% \specialrule{.4pt}{0pt}{2pt}
mDPR-X & 60.6 & 73.5 & 80.8 & 52.1 & 49.5  & 61.7 & 66.0 & 51.8 & 50.3 & 63.5 & 65.6 & 56.3& 58.9  & 73.5 & 85.9 & 76.6 & 58.2& 36.3 & 30.2 \\
mContriever-X & 86.5 & 92.0 & 95.3 & 80.6 & 78.8  & 84.0  & 93.1  & 86.0  & 82.1  & 83.7  & 89.5  & 87.7 & 86.7  & 93.3  & 96.7  & 94.3  & 85.9 & 79.3 & 68.8 \\
% \specialrule{.4pt}{2pt}{0pt}
\rowcolor{paleaqua} \multicolumn{20}{l}{\textit{Synthetic Baselines (Our work)}} \\
% \specialrule{.4pt}{0pt}{2pt}
SWIM-X (180K) & 78.9 & 89.2 & 87.8 & 72.9 & 70.0 & 76.3 & 91.6 & 75.8 & 72.5 & 74.3 & 77.6 & 76.8 & 77.9 & 87.8 & 84.9 & 92.9 & 69.9 & 72.4 & 69.3  \\
\bottomrule
\end{tabular}
}
\caption{
Experimental results for monolingual retrieval on MIRACL dev \cite{miracl}. All scores denote Recall@100; Hybrid~denotes a hybrid retriever with ranked fusion of three retrievers: mDPR, mColBERT and BM25; BM25, mDPR and Hybrid scores \cite{miracl}; Cohere-API is used as a reranker on top of 100 BM25 results \cite{kamalloo2023evaluating}. SWIM-X is fine-tuned on 180K synthetic data.}
\label{tab:miracl-results-supp}
\vspace*{-2mm}
\end{table*}

\clearpage

\clearpage

\newgeometry{scale=1}
\thispagestyle{empty}

\includepdf[pages={1-},scale=0.9, frame=true]{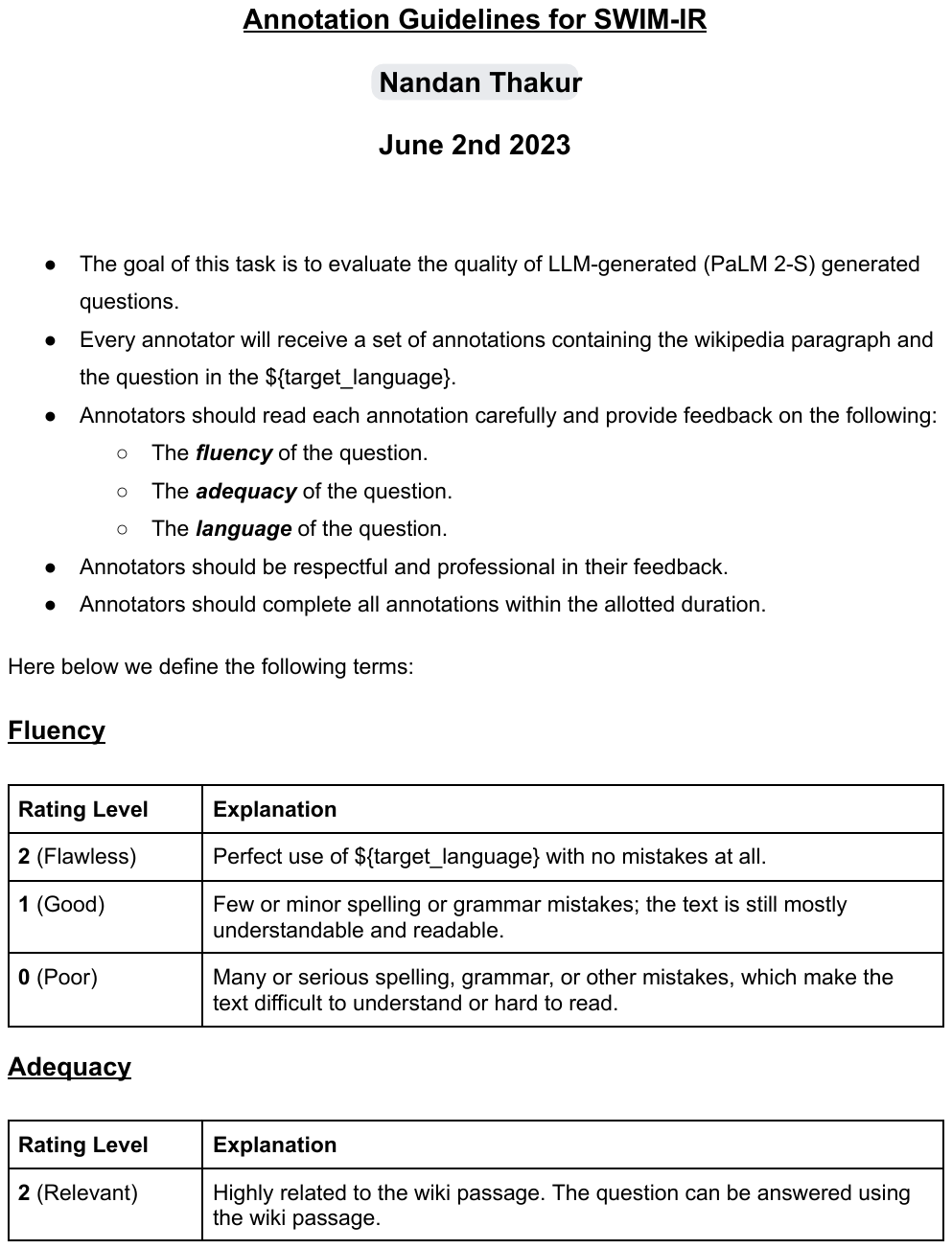}
\restoregeometry

\end{document}